\documentclass[sigconf,balance=false]{acmart}

\usepackage{float,siunitx} 
\usepackage{amsmath,amsfonts}
\usepackage{booktabs,caption,subcaption}
\usepackage{algorithmicx,setspace}
\usepackage[noend]{algpseudocode}
\usepackage{graphics,graphicx,xcolor}
\usepackage[utf8]{inputenc}
\usepackage{arydshln}
\usepackage{pst-node}
\usepackage{multido}
\usepackage{xcolor}
\usepackage{nicematrix}
\usepackage{natbib}

\usepackage{varioref,hyperref}
\hypersetup{
}
\usepackage[nameinlink,capitalize]{cleveref}

\usepackage{tikz}
\usetikzlibrary{positioning,chains,fit,shapes,calc}
\usepackage[most]{tcolorbox}
\usepackage{comment}
\usepackage[section]{placeins}

\newtheorem{definition}{\textbf{Definition}}

\newif\ifarxiv

\arxivtrue

\usepackage{pifont}
\newcommand{\cmark}{\ding{51}}%
\newcommand{\xmark}{\ding{55}}%
\usepackage{popets}

\setcopyright{popets}
\copyrightyear{YYYY}

\acmYear{YYYY}
\acmVolume{YYYY}
\acmNumber{X}
\acmDOI{XXXXXXX.XXXXXXX}
\acmISBN{}
\acmConference{Proceedings on Privacy Enhancing Technologies}
\begin{document}

\title[Towards Measuring the Traceability of Cryptocurrencies]{Towards Measuring the Traceability of Cryptocurrencies}


\author{Domokos M. Kelen}
\affiliation{%
  \institution{ELKH Institute for Computer Science and Control (SZTAKI)}
  \city{Budapest}
  \country{Hungary}}

\author{István András Seres}
\affiliation{%
  \institution{Eötvös Loránd University}
  \city{Budapest}
  \country{Hungary}}

\renewcommand{\shortauthors}{D.M. Kelen and I.A. Seres}

\begin{abstract}
    Cryptocurrencies aim to replicate physical cash in the digital realm while removing centralized and trusted intermediaries. Decentralization is achieved by the blockchain, a permanent public ledger that contains a record of every transaction. The public ledger ensures transparency, which enables public verifiability but harms untraceability, fungibility, and anonymity. In the last decade, cryptocurrencies attracted millions of users, with their total market cap reaching approximately three trillion USD at its peak. However, their anonymity guarantees are poorly understood and plagued by widespread misbeliefs. Indeed, previous notions of privacy, anonymity, and traceability for cryptocurrencies are either non-quantitative or inapplicable, e.g., computationally hard to measure.
    
    In this work, we put forward a formal framework to measure the (un)traceability and anonymity of cryptocurrencies, allowing us to quantitatively reason about the mixing characteristics of cryptocurrencies and the privacy-enhancing technologies built on top of them. Our methods apply absorbing Markov chains combined with Shannon entropy. To the best of our knowledge, our work provides the first practical, efficient, and probabilistic measure to assess the traceability of cryptocurrencies quantitatively, which also generalizes to entire cryptocurrency transaction graphs. We implement and extensively evaluate our proposed traceability measure on several cryptocurrency transaction graphs. Among other quantitative results, we find that in the studied one-week interval, the Bitcoin blockchain, on average, provided comparable but quantifiably more natural mixing than the Ethereum blockchain.
\end{abstract}

\keywords{Blockchain privacy, Untraceability, Fungibility, Anonymity.}
\maketitle
\section{Introduction} \label{sec:intro}

Traditional monetary systems and cryptocurrencies have different centralization, transparency, and openness properties. The former operates in a centralized setting, where the transaction history is not publicly available, hence, the ledger of transactions is not auditable. In stark contrast, cryptocurrencies, most notably Bitcoin~\cite{nakamoto2008bitcoin} and Ethereum~\cite{wood2014ethereum}, operate in the permissionless and decentralized setting, where all transactions are readily available for inspection in a public ledger, typically in a blockchain. The blockchain stores all valid transactions in a timestamped manner, which makes it possible to publicly verify the monetary system's state. The public nature of the ledger comes at the cost of rendering each coin in the system traceable and non-fungible, as its entire history is transparent. The privacy implications of traceability and its effects on anonymity are not hypothetical: both academic and industry efforts have been made to extract insights and intelligence from these public ledgers~\cite{chainalysis20202020,daian2019flash,meiklejohn2013fistful}. Several cryptocurrency addresses are black-listed at major cryptocurrency exchanges due to their alleged illicit activities~\cite{moser2019effective}.

The traceable nature of cryptocurrencies creates friction, inefficiencies, and ample opportunities for censorship~\cite{winzer2019temporary,wahrstatter2023blockchain,wang2023blockchain}. These major deficiencies render cryptocurrencies a subpar form of money compared to cash from a privacy point of view. The traceability of cryptocurrencies, specifically Bitcoin, was first studied by Meiklejohn et al. in 2013~\cite{meiklejohn2013fistful}. Since then, significant efforts -- both academic~\cite{ahmed2019tendrils,biryukov2019privacy,moser2017empirical,khalilov2018survey,yousaf2019tracing} and industry~\cite{barragan2021cryptocurrency,chainalysis2022chainalysis} -- have been committed to studying, understanding, and exploiting the traceability of these novel digital currencies. 

There are two sides to cryptocurrency traceability. The first line of work aims at extracting information and intelligence thanks to the traceable nature of cryptocurrencies. Typically, these works try to trace illicit funds with heuristics~\cite{moser2021resurrecting,victor2020address} and ad-hoc methods, sometimes even across multiple blockchains~\cite{hinteregger2019short}. In contrast, the second line of work aims to enhance anonymity and untraceability by employing cryptographic tools, e.g., zero-knowledge proofs~\cite{bunz2020zether,pertsev2019tornado,sasson2014zerocash} or secure multi-party computation~\cite{maxwell2013coinjoin}. To tackle these issues, besides standalone privacy-focused cryptocurrencies, such as Monero~\cite{van2013cryptonote} or Zcash~\cite{sasson2014zerocash}, several privacy-enhancing technologies (PETs) have been proposed for cryptocurrencies~\cite{bonneau2014mixcoin,heilman2017tumblebit,maxwell2013coinjoin,narayanan2017obfuscation,ruffing2014coinshuffle,valenta2015blindcoin}. Nevertheless, their quantitative benefits in terms of anonymity, (un)traceability or fungibility remain unexplained from a quantitative point of view.

In this work, we are primarily interested in what we call \emph{objective} traceability, where we measure the uncertainty that \emph{any} observer of the public ledger has about the origins of coins residing at a specific address. On the other hand, \emph{subjective} traceability measures the uncertainty of a more potent adversary about the source of some coins. In particular, when measuring subjective traceability, we allow the adversary to have external knowledge or use additional heuristics about the flow of money. This external knowledge may come from the peer-to-peer (P2P) layer~\cite{fanti2017anonymity} or any source other than the public ledger, e.g., application layer~\cite{goldfeder2017cookie}. 
In essence, objective traceability measures the maximum uncertainty given only the public ledger as an information source, while subjective traceability depends on the observer.

We believe that currently, the quantification of cryptocurrency traceability is lacking. To the best of our knowledge, there is no prior satisfactory formal quantitative framework to argue about the traceability of cryptocurrencies, e.g., Bitcoin~\cite{nakamoto2008bitcoin}, Ethereum~\cite{wood2014ethereum}, or Zcash~\cite{sasson2014zerocash}. Our main goal is to define a metric which is general enough to encompass most transaction graphs arising from real-world use cases, see~\Cref{sec:applications}. Moreover, it should be efficiently computable on real-world blockchain data. Therefore, this work provides a method to quantify cryptocurrencies' objective traceability and evaluate it on several cryptocurrency transaction graphs. Our quantitative framework for traceability allows a clear comparison of cryptocurrencies and privacy-enhancing technologies concerning their achieved untraceability benefits. We also show how one could extend our traceability framework to quantify the anonymity guarantees of cryptocurrency transaction graphs to calculate subjective measures, i.e., contingent on ``adversarial'' knowledge of traceability.

In many applications, we would like an anonymity score or degree of anonymity~\cite{diaz2002towards,serjantov2002towards}. But this is impossible on a pseudonymous blockchain, e.g., Bitcoin or Ethereum, since these blockchains do not provide strong identities. Hence, defining a useful anonymity score is futile in our context. On the other hand, anonymity and traceability are closely related. Traceability is a good proxy for assessing the anonymity of \emph{coins} or \emph{outputs}. Further evidence for this lies in the fact that most privacy-enhancing technologies explicitly focus on making it harder to trace the source of funds on transaction graphs. 

Despite this relation, we want to emphasize that traceability is not a privacy or anonymity metric. In a typical (information-theoretic) anonymity metric~\cite{diaz2002towards,serjantov2002towards}, the adversary observes a communication system and outputs a probability distribution of the possible targets in an anonymity set. The entropy of this distribution captures the adversary's uncertainty about the target. Therefore, the anonymity of a communication system depends on adversarial capabilities and background knowledge. In contrast, we wish to provide an absolute traceability metric that does not depend on external components but solely on the public ledger containing all transactions. Informally, an untraceability metric should measure how ``unique'' unspent coins are in a cryptocurrency.

The uniqueness of a coin is determined by its transaction history, which we model in Section~\ref{sec:theory}. The two main applications that motivate the introduction of an untraceability metric are as follows.
\begin{itemize}
    \item \textbf{Privacy-focused cryptocurrency wallet.} A user typically owns several coins (or addresses) in a cryptocurrency wallet. However, the coins owned by the user might have differing levels of traceability. A traceability measure could help users select the coin with sufficient traceability/anonymity required to perform a privacy-critical transaction as described in~\Cref{sec:applications}.  
    \item \textbf{Benchmarking cryptocurrency and PET designs.} A traceability and anonymity metric allows us to compare the efficacy of various privacy-preserving cryptocurrency designs quantitatively, e.g., Zcash~\cite{sasson2014zerocash}, Monero~\cite{van2013cryptonote} or Dash~\cite{duffield2015dash}. Similarly, one could quantitatively study the traceability and anonymity guarantees of privacy-enhancing overlays, e.g., mixers~\cite{bonneau2014mixcoin,tran2018obscuro}, stealth addresses~\cite{courtois2017stealth}, tumblers~\cite{heilman2016tumblebit,meiklejohn2018mobius,pertsev2019tornado,ruffing2014coinshuffle,seres2019mixeth}, etc., built on top of cryptocurrencies.
\end{itemize}

In this work, we provide the following contributions.
\begin{itemize}
    \item \textbf{Quantitative theoretical framework for cryptocurrency (un)traceability.} 
    We propose a model to quantify the traceability of cryptocurrencies. Our traceability metric can be considered a generalization of the information-theoretic anonymity metric by Diaz et al.~\cite{diaz2002towards} for cryptocurrencies and transaction graphs.
    \item \textbf{Empirical evaluation on major blockchains.}
We extensively evaluate our proposed traceability metric on major cryptocurrencies,
such as Bitcoin, Ethereum, Zcash, and some ERC-20 tokens. For the first
time, we are able to assess the traceability of crypto assets quantitatively.
    \item \textbf{Open-source implementation.} The implementation of our traceability metric is open-sourced and available for reproduction upon publication.
\end{itemize}

The rest of this paper is organized as follows. In~\Cref{sec:model}, we provide our system and transaction graph models and define our proposed untreaceability score. In~\Cref{sec:theory}, we apply our model to cryptocurrencies. 
We detail a few applications of our metric in~\Cref{sec:applications}. In~\Cref{sec:evaluation}, we evaluate our proposed traceability measure on the public ledgers of some cryptocurrencies. We discuss limitations and possible extensions in~\Cref{sec:discussion}. In~\Cref{sec:relatedwork}, we review previous attempts to quantify the privacy provisions of cryptocurrencies. Finally, we conclude our work in~\Cref{sec:conclusion} with open research directions.

\section{Untraceability} \label{sec:model}
First, we describe our system model. Afterwards, we introduce our transaction graph model, which we subsequently convert into an absorbing Markov chain to define a metric of untraceability.

\subsection{System Model}
In our model, we distinguish between three system components: \emph{users}, the public \emph{ledger}, and \emph{adversaries}.

\begin{itemize}
    \item \textbf{Users} issue transactions. We assume users can issue transactions in a privacy-preserving manner. In other words, we do not model privacy leakages on the network and application layers and solely focus on the untraceability achieved by the issued and recorded transactions.
    \item \textbf{Ledger} is an immutable, public database recording faithfully every transaction users perform. The ledger is known to every participant of the system. We model the ledger as a weighted, directed graph, see~\Cref{sec:matmodel}.
    \item \textbf{Adversary} aims to establish the sources of each coin with respect to a certain ``source'' set when it has a particular ``knowledge'' set (both of these terms are defined precisely later). We provide an untraceability metric with respect to a given ``source'' set from the adversary's perspective, i.e., when the adversary relies on a specific knowledge set. The adversary might possess external information sources, e.g., social media, application layer data, etc., and tools, e.g., logging broadcast transaction IP addresses, to increase its tracing capabilities. While the adversary model in this paper does not utilize such information sources, our model can naturally incorporate them (in the applied transition matrix) whenever they are available, see~\Cref{sec:discussion}.
\end{itemize}

\subsection{Transaction Graph Model}\label{sec:matmodel}
We define a general transaction graph model that can be used to calculate untraceability, independent of the exact implementation details of cryptocurrency transaction networks. We discuss the transformation of real-world cryptocurrency networks into transaction graphs in~\Cref{sec:theory}.

A transaction graph $G$ is a weighted, directed graph $\overrightarrow{G}=(V,E,a)$, where $V$ consists of addresses, $E\subseteq V\times V$ is a set of ordered pairs that denotes transactions, while $a:V\times V\xrightarrow{}\mathbb{N}$ assigns the currency amounts $a((v_1,v_2))$ transferred from $v_1$ to $v_2$. Note that $a((v_1,v_2))=0$ if $(v_1,v_2)\notin E$. The definition $\mathit{Im}(a)=\mathbb{N}$ is not a restriction since we can consider the smallest denomination of the currency, for example, satoshi for Bitcoin or wei for Ether. We set no restrictions about a node's incoming and outgoing amounts: nodes with incoming surplus become the sinks, while nodes with outgoing surplus become the sources of the transaction graph.

Looking ahead, we think of $V$ as the set of all entities in the ledger: addresses of users or smart contracts, as well as time-snapshots of addresses in some cases, see~\Cref{sec:temporaltxgraphs}. Multiple details about converting ledgers into transaction graphs require special attention, e.g., multiple transactions between two addresses or the local graph structures resulting from the use of UTXO-based currencies. We detail these transformations in~\Cref{sec:theory}.

\subsection{Desiderata}\label{sec:desiderata}
We want to provide an untraceability score $U_{S,G,\mathcal{A}}(\cdot)$ defined on the nodes $V$ of the transaction graph $G$ using adversarial knowledge $\mathcal{A}$, i.e., $U_{S,G,\mathcal{A}}:V\rightarrow\mathbb{R}^{+}_{0}$ with respect to a source set $\emptyset\neq S\subset V$. Note that in the following, we will omit from the subscript the set of source nodes $S$ to which we trace the money back, the transaction graph $G$, or the adversary $\mathcal{A}$ whenever they are irrelevant or obvious from the context to avoid notational clutter.

\begin{figure}
    \centering
    \vspace{-5em}
    \includegraphics[width=\linewidth,clip]{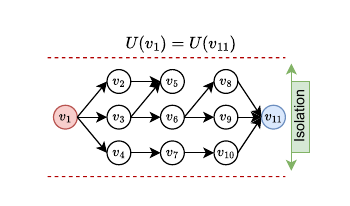}
    \vspace{-3em}
    \caption{Illustrating the requirement of ``tamper resistance'' for our proposed metric, see~\Cref{sec:desiderata}. Intuitively, the traceability metric for $v_1$ and $v_{11}$ should be the same since no third-party coins are mixed into $v_1$.}
    \label{fig:tamper-resistance-explanation}
\end{figure}

Next, we outline the requirements we believe a traceability quantification should possess.
\begin{itemize}
    \item \textbf{Range}. We want to have a positively or negatively oriented range, where one extreme of the spectrum signifies total untraceability (i.e., k-anonymity), while the other extreme is total traceability (i.e., exact identification of source).
    \item \textbf{Fine-grained}. Instead of a binary metric, we want the metric to reflect the \emph{degree} of traceability, also capable of capturing constructs between total traceability and total untraceability.
    \item \textbf{Decay}. We want the traceability value to naturally decrease with time: as a source gets mixed with other sources, its traceability should decrease. Formally, we want the maximum traceability to decrease (or remain constant) whenever edges are added to a subgraph.
    \item \textbf{Tamper-resistant}. We want the measure to be such that without the use of third-party funds (i.e., in isolation), the traceability cannot be decreased. For an illustrative example, see~\Cref{fig:tamper-resistance-explanation}.
    
    \item\textbf{Subjective}. The metric should be able to incorporate background knowledge (i.e., private information) to yield a subjective measure of traceability.
    \item \textbf{Efficient}. It should be computationally feasible to calculate in practice, i.e., for graphs with millions of nodes and edges.
\end{itemize}

We review and evaluate previous approaches in~\Cref{sec:relatedwork} and~\Cref{table:approaches_qualitative_comparison}, whether they satisfy our desiderata. For instance, the previously proposed Boltzmann score does not satisfy our efficiency requirement as it reduces to the NP-hard subset sum problem. Similarly, the system's anonymity metric~\cite{edman2007combinatorial,gierlichs2008revisiting} reduces to computing the permanent of the transaction graph's adjacency matrix, again an NP-hard problem. In~\Cref{sec:defining_untraceability}, we define our traceability metric and argue why it satisfies our desiderata above.

\subsection{Tracing Money Backwards} 
\label{sec:tx_graph_to_markov_chain}
A transaction graph $G$ has many source nodes. The source nodes correspond to transactions that spend money they did not receive. These can either be coinbase transactions that mint money into existence or nodes with an initial non-zero balance in the case of a time snapshot of a transaction graph. Let us denote the set of source nodes in the transaction graph as $\mathsf{S}$, i.e., $\mathsf{S}\subseteq V$. Similarly, many nodes act as sinks in the transaction graph, i.e., they do not spend the money they receive. To trace the origin of each output residing in sink nodes to source nodes, we reverse the direction of the edges and start random walks on the reversed edges.

\begin{figure}[ht!]
\begin{subfigure}[t]{.48\textwidth}
    \centering
    \includegraphics[width=0.5\linewidth,clip]{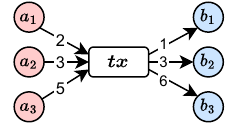}
    \caption{A simple transaction in a UTXO-based cryptocurrency. Each input $(a_i)^{3}_{i=1}$ and output $(b_i)^{3}_{i=1}$ are connected through a $\mathit{tx}$ node. The weights on the edges represent the value of the inputs and outputs used in the transaction.}
    \label{fig:utxo_tx_graph}
\end{subfigure}\hfill%
\begin{subfigure}[t]{.48\textwidth}
       \centering
    \includegraphics[width=0.5\linewidth,clip]{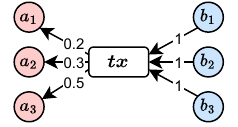}
    \caption{The transition probability graph of the example transaction from~\Cref{fig:utxo_tx_graph}. Note that edges are reversed, i.e., the random walk proceeds from outputs towards the transaction's inputs. Numbers on the edges represent transition probabilities according to the transition matrix, cf.~\Cref{eq:transmatrixdef}.}
    \label{fig_markov_chain} 
\end{subfigure}
\caption{Transition probabilities in the Markov chain.}
\label{fig:graph_to_markov}
    
\end{figure}

While tracing can naturally be performed along a backwards topological order if the transaction graph has no loops, for the general case, we will have to define random walks and Markov chains as follows. 
First, we extend the set of nodes of the graph $G$ by creating an auxiliary node $v'\in V'$ for each source node $v$ in $G$ with a corresponding edge $(v',v)\in E$ and weight $a((v',v))$. This edge represents an imaginary transaction that accounts for the outgoing surplus of $v$. This way, only the newly created nodes $V'$ act as sources, and all of the original nodes $V$ either act as transient nodes or sinks. More importantly, the source nodes of the modified graph are guaranteed to have no incoming edges. The newly created nodes $V'$ together with the original nodes $V$ of $G$ form the states $U=V\cup V'$ of the Markov chain $G^*$ with edge directions reversed.

When reversing the edges of $G$, the nodes in $V'$ become absorbers of the Markov chain. Intuitively, this is done by adding self-loops to these nodes so that a random walk cannot escape once it reaches an absorber node. Ultimately, we are interested in computing the distribution of absorbing probabilities over the absorbers when starting random walks from sinks of the original graph $G$. We denote the transition probability matrix as $\mathbf{P}$, where $\mathbf{P}_{i,j}$ denotes the transition probability from state $u_i$ to $u_j$ in the Markov chain. We define
\begin{align} \label{eq:transmatrixdef}
    \mathbf{P}_{i,j}&\stackrel{def}{=}
\begin{cases}
    1, & \text{for absorber self-loops, i.e., $u_i=u_j\in V'$}\\
	\frac{a((u_j,u_i))}{\sigma(u_i)}, & \text{for reversed edges, i.e., $u_i\in V$, $a((u_j,u_i))>0$}\\
    0 & \text{otherwise,}
 \end{cases}
\end{align}
where $\sigma(u_i) =\sum_{u_k\in U}a((u_k,u_i))$, i.e., the sum of the incoming transactions to $u_i$. Intuitively, this means that each output of a node traces back to all inputs of the node, proportionally to the transferred amounts. For an example of this calculation, see~\Cref{fig:graph_to_markov}.
The Markov chain defined this way is an absorbing Markov chain, meaning that each walk from a sink of $G$ must eventually reach an absorbing state with probability $1$. For formal proof of this claim, see~\Cref{sec:convergence}.

\subsubsection{Other policies}
While we consider the transaction graph model to be given, tracking money backwards is highly subjective and depends on the adversary's goals and background knowledge. The random walk's transition matrix $\mathbf{P}$ can capture these different, subjective background knowledge. Various adversaries with varying background knowledge can define their own subjective untraceability score using their ``custom'' transition matrices.

To illustrate the flexibility of our proposed formal model, we mention an alternative transition matrix inspired by the risk-scoring models of Möser, Böhme, and Breuker~\cite{moser2014towards}. In particular, we define the transition matrices $,\mathbf{P}^{\textsf{pois}}$ corresponding to the \emph{poison risk scoring policy}. The poison policy defines a set of tainted addresses $T\subseteq V$. If a transaction has even a single tainted input address, then all output addresses become tainted as well. The poison policy can be formulated as a transition matrix as follows.

\begin{align} \label{eq:transmatrixPoisdef}
    \mathbf{P}^{\textsf{pois}}_{i,j}&\stackrel{def}{=}
\begin{cases}
    1, & \text{for absorber self-loops, i.e., $u_i=u_j\in V'$}\\
	\frac{1}{d(u_j)}, & \text{for reversed edges, $u_j\in V$, for $d(u_j))>0$}\\
    0 & \text{otherwise,}
 \end{cases}
\end{align}
where $d(u_j)$ is the number of tainted neighbor nodes  of $u_j$ in $V$. Similarly, other risk scoring or black listing policies can be easily incorporated into our formal model by adjusting the Markovian random walk's transition matrix $\mathbf{P}$ accordingly.

In the rest of this work, we will apply and evaluate on several cryptocurrency transaction graphs the transition matrix defined in~\Cref{eq:transmatrixdef}, cf.~\Cref{sec:evaluation}.

To calculate the absorbing probabilities, we follow the outline of Kemény and Snell~\cite{kemeny1983finite}. Let $G^{*}$ have $t=|V|$ transient states and $r=|V'|$ absorbing states. First, we rearrange the entries of $\mathbf{P}$ as follows. Let
\begin{equation}\label{eq:markov_normalform}
    \mathbf{P}=
  \begin{bmatrix}
    \mathbf{Q} & \mathbf{R} \\
    \mathbf{0} & \mathbf{I}_r
  \end{bmatrix},
\end{equation}
where $\mathbf{Q}$ is a $t$-by-$t$ matrix representing transition probabilities from $V$ to $V$, $\mathbf{R}$ is a $t$-by-$r$ matrix representing transition probabilities from $V$ to $V'$ and $\mathbf{I}_r$ is the $r$-by-$r$ identity matrix. The probability of transitioning from $v_i$ to $v_j$ in exactly $k$ steps is the $(i,j)$ entry of $\mathbf{Q}^{k}$. The probabilities of being absorbed in state $v_j\in V'$ starting from transient state $v_i\in V$ can be obtained as the $(i,j)$ element of the $t\times r$ absorption probability matrix $\mathbf{B}$, which is computed as
\begin{equation}
    \mathbf{B}=\lim_{n\xrightarrow{}\infty}\mathbf{P}^{n}\textrm{.}
\end{equation}
Each row $\mathbf{B}_i$ of the resulting matrix $\mathbf{B}$ represents a probability distribution over the absorbing nodes $V'$ when starting from node $v_i\in V$, i.e., $ \forall i,j: 0\leq\mathbf{B}_{i,j}\leq 1 $ and $\sum_{j}\mathbf{B}_{i,j}=1$.

\subsection{Defining untraceability}\label{sec:defining_untraceability}

Having the necessary mathematical framework in place, we can now continue by defining a quantitative measure of traceability. However, as it turns out, an \emph{un}traceability score is more natural, as it is analogous to the degree of anonymity~\cite{diaz2002towards}.

\begin{definition}[\emph{\textbf{Untraceability score}}]
\label{def:fungibility}
The untraceability score is defined as \emph{the amount of information the adversary lacks} (in bits) for achieving total certainty in tracing the money. More formally, the untraceability $U_{\scriptscriptstyle\mathsf{S},G,\mathcal{A}}(v_i)$ of a sink $v_i$ in the transaction graph $G$ with respect to a source set $\mathsf{S}$ and adversary $\mathcal{A}$ is defined as
\begin{equation}\label{eq:untraceability_definition}
    U_{\scriptscriptstyle\mathsf{S},G,\mathcal{A}}(v_i):=H(\mathbf{B}_i), 
\end{equation}
where $H(\mathbf{B}_i)$ is the Shannon entropy (in bits) of the probability distribution $\mathbf{B}_{i}$. Often, we omit $\mathsf{S},G,\mathcal{A}$ from the subscript of $U_{\scriptscriptstyle\mathsf{S},G,\mathcal{A}}(v_i)$ if they are clear from the context or irrelevant.
\end{definition}

\begin{figure*}[ptbh]
    \centering

    \begin{subfigure}[t]{0.43\textwidth}
        \centering
        \includegraphics[width=0.9\linewidth,clip]{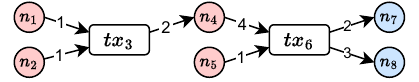}
        \caption*{Transaction graph $G$}
    \end{subfigure}
    \begin{subfigure}[t]{0.55\textwidth}
        \centering
        \includegraphics[width=0.9\linewidth,clip]{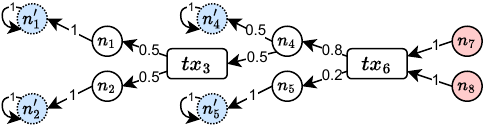}
        \caption*{Corresponding absorbing Markov chain}
    \end{subfigure}

    \vspace{1em}
    \begin{subfigure}[t]{\textwidth}
        \centering
        \setlength{\tabcolsep}{0.3em} 
        \begin{tabular}{ccc}
            $\begin{bNiceMatrix}[first-row,first-col]
            &\scriptstyle n_1 & \scriptstyle n_2 & \scriptstyle tx_3 & \scriptstyle n_4 & \scriptstyle n_5 & \scriptstyle tx_6 & \scriptstyle n_7 & \scriptstyle n_8 \\
            \scriptstyle n_1&{\color{gray} 0} & {\color{gray} 0} & 1 & {\color{gray} 0} & {\color{gray} 0} & {\color{gray} 0} & {\color{gray} 0} & {\color{gray} 0} \\
            \scriptstyle n_2&{\color{gray} 0} & {\color{gray} 0} & 1 & {\color{gray} 0} & {\color{gray} 0} & {\color{gray} 0} & {\color{gray} 0} & {\color{gray} 0} \\
            \scriptstyle tx_3&{\color{gray} 0} & {\color{gray} 0} & {\color{gray} 0} & 2 & {\color{gray} 0} & {\color{gray} 0} & {\color{gray} 0} & {\color{gray} 0} \\
            \scriptstyle n_4&{\color{gray} 0} & {\color{gray} 0} & {\color{gray} 0} & {\color{gray} 0} & {\color{gray} 0} & 4 & {\color{gray} 0} & {\color{gray} 0} \\
            \scriptstyle n_5&{\color{gray} 0} & {\color{gray} 0} & {\color{gray} 0} & {\color{gray} 0} & {\color{gray} 0} & 1 & {\color{gray} 0} & {\color{gray} 0} \\
            \scriptstyle tx_6&{\color{gray} 0} & {\color{gray} 0} & {\color{gray} 0} & {\color{gray} 0} & {\color{gray} 0} & {\color{gray} 0} & 2 & 3 \\
            \scriptstyle n_7&{\color{gray} 0} & {\color{gray} 0} & {\color{gray} 0} & {\color{gray} 0} & {\color{gray} 0} & {\color{gray} 0} & {\color{gray} 0} & {\color{gray} 0} \\
            \scriptstyle n_8&{\color{gray} 0} & {\color{gray} 0} & {\color{gray} 0} & {\color{gray} 0} & {\color{gray} 0} & {\color{gray} 0} & {\color{gray} 0} & {\color{gray} 0}
            \end{bNiceMatrix}$ &
            $\begin{bNiceMatrix}[first-row]
            \scriptstyle b \\
            1.0 \\
            1.0 \\
            {\color{gray} 0}\\
            2.0\\
            1.0 \\
            {\color{gray} 0}\\
            {\color{gray} 0}\\
            {\color{gray} 0}\\
            \end{bNiceMatrix}$\vspace{0.5em}\\
            \hphantom{$tx_3$}Adjacency matrix & Initial balances
        \end{tabular}
    \end{subfigure}

    \vspace{1em}
    \begin{subfigure}[t]{\textwidth}
        \centering
        \begin{tabular}{cccc}
        $\begin{bNiceMatrix}[first-row,first-col]
            &\scriptstyle n_1 & \scriptstyle n_2 & \scriptstyle tx_3 & \scriptstyle n_4 & \scriptstyle n_5 & \scriptstyle tx_6 & \scriptstyle n_7 & \scriptstyle n_8 \\
            \scriptstyle n_1&{\color{gray} 0} & {\color{gray} 0} & {\color{gray} 0} & {\color{gray} 0} & {\color{gray} 0} & {\color{gray} 0} & {\color{gray} 0} & {\color{gray} 0} \\
            \scriptstyle n_2&{\color{gray} 0} & {\color{gray} 0} & {\color{gray} 0} & {\color{gray} 0} & {\color{gray} 0} & {\color{gray} 0} & {\color{gray} 0} & {\color{gray} 0} \\
            \scriptstyle tx_3&0.5 & 0.5 & {\color{gray} 0} & {\color{gray} 0} & {\color{gray} 0} & {\color{gray} 0} & {\color{gray} 0} & {\color{gray} 0} \\
            \scriptstyle n_4&{\color{gray} 0} & {\color{gray} 0} & 0.5 & {\color{gray} 0} & {\color{gray} 0} & {\color{gray} 0} & {\color{gray} 0} & {\color{gray} 0} \\
            \scriptstyle n_5&{\color{gray} 0} & {\color{gray} 0} & {\color{gray} 0} & {\color{gray} 0} & {\color{gray} 0} & {\color{gray} 0} & {\color{gray} 0} & {\color{gray} 0} \\
            \scriptstyle tx_6&{\color{gray} 0} & {\color{gray} 0} & {\color{gray} 0} & 0.8 & 0.2 & {\color{gray} 0} & {\color{gray} 0} & {\color{gray} 0} \\
            \scriptstyle n_7&{\color{gray} 0} & {\color{gray} 0} & {\color{gray} 0} & {\color{gray} 0} & {\color{gray} 0} & 1 & {\color{gray} 0} & {\color{gray} 0} \\
            \scriptstyle n_8&{\color{gray} 0} & {\color{gray} 0} & {\color{gray} 0} & {\color{gray} 0} & {\color{gray} 0} & 1 & {\color{gray} 0} & {\color{gray} 0} 
            \end{bNiceMatrix}$&
            $\begin{bNiceMatrix}[first-row]
            \scriptstyle n_1' & \scriptstyle n_2' & \scriptstyle n_4' & \scriptstyle  n_5' \\
            1 & {\color{gray} 0} & {\color{gray} 0} & {\color{gray} 0} \\
            {\color{gray} 0} & 1 & {\color{gray} 0}  & {\color{gray} 0}\\
            {\color{gray} 0} & {\color{gray} 0} & {\color{gray} 0}  & {\color{gray} 0}\\
            {\color{gray} 0} & {\color{gray} 0} & {\color{gray} 0} & 0.5\\
            {\color{gray} 0} & {\color{gray} 0} & 1 & {\color{gray} 0} \\
            {\color{gray} 0} & {\color{gray} 0} & {\color{gray} 0} & {\color{gray} 0} \\
            {\color{gray} 0} & {\color{gray} 0} & {\color{gray} 0} & {\color{gray} 0} \\
            {\color{gray} 0} & {\color{gray} 0} & {\color{gray} 0} & {\color{gray} 0} 
            \end{bNiceMatrix}$&
            $\begin{bNiceMatrix}[first-row]
            \scriptstyle n_1' & \scriptstyle n_2' & \scriptstyle n_4' & \scriptstyle n_5'\\
            1.0  & {\color{gray} 0}  & {\color{gray} 0}  & {\color{gray} 0}\\
            {\color{gray} 0}   & 1.0   & {\color{gray} 0}  & {\color{gray} 0}\\
            0.5 & 0.5 & {\color{gray} 0}  & {\color{gray} 0}\\
            0.25 & 0.25 & 0.5 & {\color{gray} 0}\\
            {\color{gray} 0} & {\color{gray} 0} & 1.0  & {\color{gray} 0}\\
            0.2 & 0.2 & 0.4 & 0.2\\
            0.2 & 0.2 & 0.4 & 0.2\\
            0.2 & 0.2 & 0.4 & 0.2
            \end{bNiceMatrix}$ &
            $\begin{bNiceMatrix}[first-row]
                \scriptstyle U(v)\\
                {\color{gray} -}\\
                {\color{gray} -}\\
                {\color{gray} -}\\
                {\color{gray} -}\\
                {\color{gray} -}\\
                {\color{gray} -}\\
                1.92\\
                1.92
            \end{bNiceMatrix}$\vspace{0.5em}\\
            \hphantom{$tx_3$}$\mathbf{Q}$&$\mathbf{R}$ & $\mathbf{B}$ & Untraceability
        \end{tabular}
    \end{subfigure}
    \caption{A simple example summarizing our method to obtain an untraceability score on
    a transaction graph $G$.
    Red nodes are sources, blue nodes are sinks. While outputs of transactions cannot be sources in a regular UTXO-based graph, the initial balance of node $n_4$ is possible when merging different UTXOs belonging to the same public key.
    We add an auxiliary absorber node in the corresponding Markov chain for each source in $G$, i.e., each node with an initial balance. The transition matrix of the Markov chain is split into $\mathbf{Q}$ and $\mathbf{R}$ as described in~\Cref{eq:markov_normalform} to calculate the absorbing probabilities $\mathbf{B}=(\mathbf{I}-\mathbf{Q})^{-1}\mathbf{R}$. The untraceability score is defined as the Shannon entropy of the distribution of the absorbing probabilities for each sink in the original transaction graph $G$.}
    \label{fig:txgraphtransformation}
\end{figure*}

\begin{figure*}[htbp]
    \centering

    \begin{subfigure}[t]{0.25\textwidth}
        \centering
        \includegraphics[width=\linewidth,trim={0 0 0 0},clip]{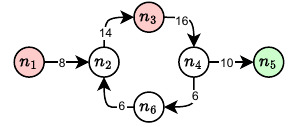}
        \caption*{Transaction graph $G$}
    \end{subfigure}
    \begin{subfigure}[t]{0.30\textwidth}
        \centering
        \includegraphics[width=\linewidth,clip]{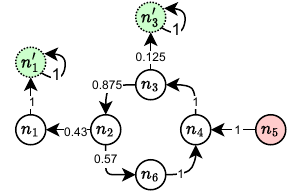}
        \caption*{Corresponding absorbing Markov chain}
    \end{subfigure}
    \begin{subfigure}[t]{0.35\textwidth}
        \centering
        \vspace{-8.4em}
        \setlength{\tabcolsep}{0.3em} 
        \begin{tabular}{cc}
            $\begin{bNiceMatrix}[first-row,first-col]
            &\scriptstyle n_1 & \scriptstyle n_2 & \scriptstyle n_3 & \scriptstyle n_4 & \scriptstyle n_5 & \scriptstyle n_6\\
            \scriptstyle n_1&{\color{gray} 0} & 8 & {\color{gray} 0} & {\color{gray} 0} & {\color{gray} 0} & {\color{gray} 0}\\
            \scriptstyle n_2&{\color{gray} 0} & {\color{gray} 0} & 14 & {\color{gray} 0} & {\color{gray} 0} & {\color{gray} 0}\\
            \scriptstyle n_3&{\color{gray} 0} & {\color{gray} 0} & {\color{gray} 0} & 16 & {\color{gray} 0} & {\color{gray} 0}\\
            \scriptstyle n_4&{\color{gray} 0} & {\color{gray} 0} & {\color{gray} 0} & {\color{gray} 0} & 10 & 6\\
            \scriptstyle n_5&{\color{gray} 0} & {\color{gray} 0} & {\color{gray} 0} & {\color{gray} 0} & {\color{gray} 0} & {\color{gray} 0}\\
            \scriptstyle n_6&{\color{gray} 0} & 6 & {\color{gray} 0} & {\color{gray} 0} & {\color{gray} 0} & {\color{gray} 0}
            \end{bNiceMatrix}$ &
            $\begin{bNiceMatrix}[first-row]
            \scriptstyle b \\
            8.0 \\
            {\color{gray} 0} \\
            2.0\\
            {\color{gray} 0}\\
            {\color{gray} 0} \\
            {\color{gray} 0}
            \end{bNiceMatrix}$\vspace{0.5em}\\
            \hphantom{$tx_3$}Adjacency matrix & Initial balances
        \end{tabular}
    \end{subfigure}

    \vspace{1em}
    \begin{subfigure}[t]{\textwidth}
        \centering
        \begin{tabular}{cccc}
        $\begin{bNiceMatrix}[first-row,first-col]
            &\scriptstyle n_1 & \scriptstyle n_2 & \scriptstyle n_3 & \scriptstyle n_4 & \scriptstyle n_5 & \scriptstyle n_6\\
            \scriptstyle n_1&{\color{gray} 0} & {\color{gray} 0} & {\color{gray} 0} & {\color{gray} 0} & {\color{gray} 0} & {\color{gray} 0}\\
            \scriptstyle n_2&0.43 & {\color{gray} 0} & {\color{gray} 0} & {\color{gray} 0} & {\color{gray} 0} & 0.57\\
            \scriptstyle n_3& {\color{gray} 0} & 0.875 & {\color{gray} 0} & {\color{gray} 0} & {\color{gray} 0} & {\color{gray} 0}\\
            \scriptstyle n_4&{\color{gray} 0} & {\color{gray} 0} & 1 & {\color{gray} 0} & {\color{gray} 0} & {\color{gray} 0}\\
            \scriptstyle n_5&{\color{gray} 0} & {\color{gray} 0} & {\color{gray} 0} & 1 & {\color{gray} 0} & {\color{gray} 0}\\
            \scriptstyle n_6&{\color{gray} 0} & {\color{gray} 0} & {\color{gray} 0} & 1 & {\color{gray} 0} & {\color{gray} 0}
            \end{bNiceMatrix}$&
            $\begin{bNiceMatrix}[first-row]
            \scriptstyle n_1' & \scriptstyle n_3'\\
            1 & {\color{gray} 0} \\
            {\color{gray} 0} & {\color{gray} 0}\\
            {\color{gray} 0} & 0.125\\
            {\color{gray} 0} & {\color{gray} 0}\\
            {\color{gray} 0} & {\color{gray} 0} \\
            {\color{gray} 0} & {\color{gray} 0}
            \end{bNiceMatrix}$&
            $\begin{bNiceMatrix}[first-row]
            \scriptstyle n_1' & \scriptstyle n_3'\\
            1.0  & 0\\
            0.857   & 0.143\\
            0.75 & 0.25\\
            0.75 & 0.25\\
            0.75 & 0.25\\
            0.75 & 0.25\\
            \end{bNiceMatrix}$ &
            $\begin{bNiceMatrix}[first-row]
                \scriptstyle U(v)\\
                {\color{gray} -}\\
                {\color{gray} -}\\
                {\color{gray} -}\\
                {\color{gray} -}\\
                0.81\\
                {\color{gray} -}
            \end{bNiceMatrix}$\vspace{0.5em}\\
            \hphantom{$tx_3$}$\mathbf{Q}$&$\mathbf{R}$ & $\mathbf{B}$ & Untraceability
        \end{tabular}
    \end{subfigure}
    \caption{An illustrative example calculation of untraceability where the transaction graph $G$ includes a directed cycle. Note that directed cycles in transaction graphs are only possible in the account-based model or, alternatively, in a UTXO-based model if one is able to merge multiple UTXOs corresponding to the same public key.}
    \label{fig:cycle_example}
\end{figure*}

Intuitively, $U_{\scriptscriptstyle\mathsf{S},G,\mathcal{A}}(v_i)$ measures the uncertainty about the origins of money residing at a sink node $v_i$ of the transaction graph $G$ with respect to the sources $\mathsf{S}$ and knowledge $\mathcal{A}$. One might define untraceability with other entropy functions $H(\cdot)$ such as min-entropy or the many variants of Rényi-entropy~\cite{renyi1961measures}. However, we stick to the Shannon entropy to remain consistent with prior work that uses Shannon entropy to measure the degree of anonymity in various mix networks~\cite{diaz2002towards,serjantov2002towards}.  For a full example of calculating untraceability on a transaction graph, see~\Cref{fig:txgraphtransformation}. For an illustrative example calculation where $G$ includes a directed cycle, see~\Cref{fig:cycle_example}.

Next, we argue informally why our introduced untraceability measure satisfies our previously introduced desiderata. The properties \textbf{range} and \textbf{fine-grained} are obviously satisfied. For \textbf{subjective}, see~\Cref{sec:applications,sec:discussion}: the definition of the transaction graph that one uses can easily incorporate epistemic knowledge while only using public information assumes the public as the subject. \textbf{Efficiency} and computation are discussed in Section~\ref{sec:computational}.

\textbf{Tamper-resistance} is true due to the properties of the Markov chain, as any isolated subgraph with given inputs and outputs can be contracted into a single point with the same inputs and outputs, without changing the probability values calculated for any output of the graph. Finally, \textbf{decay} holds, as mixing two sources always results in the traceability decreasing for the output with respect to the maximum traceability of the outputs, while splitting a source does not affect traceability. The same, applied to \emph{un}traceability, can be expressed formally as $\forall v_1,v_2\in V$ if $\exists (v_1,v_3),(v_2,v_3)\in E:  U(v_3)\geq\min(U(v_1),U(v_2))$.


\subsection{Computational considerations}
\label{sec:computational} Another characterization of $\mathbf{B}$ yields a more efficient way of computing our proposed score: 
\begin{equation}  \mathbf{B}=\lim_{n\xrightarrow{}\infty}\mathbf{P}^{n}=\mathbf{N}\mathbf{R}\textrm{,}
\end{equation}
where $\mathbf{N}$ is called the fundamental matrix. It can be shown that
\begin{equation}\label{eq:fundamental_matrix}
    \mathbf{N}=\sum^{\infty}_{k=0} \mathbf{Q}^k=(\mathbf{I}_t-\mathbf{Q})^{-1}\textrm{.}
\end{equation}
We are also interested in computing \emph{the length of the random walks} before they are absorbed in an absorber node. The expected number of steps before being absorbed when starting from transient state $v_i$ is the $i$th entry of the vector $\mathbf{t}=\mathbf{N}\mathbf{1}$, where $\mathbf{1}$ is a length-$t$ column vector whose entries are all $1$. 
Inverting the matrix $(\mathbf{I}_t-\mathbf{Q})$ is computationally infeasible in practice if the transition graph grows too large. In our large-scale evaluations, we approximate $\mathbf{B}$ instead as
\begin{gather}\label{eq:Bapprox}
    \mathbf{B}=\left(\sum^{\infty}_{k=0} \mathbf{Q}^k\right)\mathbf{R} \approx \sum^{k_{\text{max}}}_{k=0} (\mathbf{Q}^k \mathbf{R})=\mathbf{R}+\mathbf{QR}+\mathbf{Q}^2\mathbf{R}+\ldots+\mathbf{Q}^{k_{\text{max}}}\mathbf{R}\textrm{.}
\end{gather}
The formula at the right-hand side of~\Cref{eq:Bapprox} can be calculated by repeatedly multiplying a variable by $\mathbf{Q}$ and adding the value to an accumulator variable at each step. Instead of using a fixed constant $k_{\text{max}}$, we stop the calculation when
\begin{equation}\label{eq:kthreshold}
    \|\mathbf{Q}^k\mathbf{R}\|_{\text{max}}\leq \delta_{\text{threshold}}
\end{equation} is reached, which is guaranteed to happen as $\lim\limits_{k\rightarrow \infty} \|\mathbf{Q}^k\mathbf{R}\|_{\text{max}} = 0$, for a formal proof see~\Cref{sec:convergence}. An advantage of approximating $\mathbf{B}$ this way is that the right-hand side of~\Cref{eq:Bapprox} can be computed in parallel by dividing $\mathbf{R}$ column-wise and processing each piece separately. We discuss the choice of $\delta_{\text{threshold}}$ in our large-scale evaluation in~\Cref{sec:evaluation}.

\subsection{Markov Chain Convergence}\label{sec:convergence}
We prove that the random walks starting from sinks of the transaction graph indeed get absorbed with probability 1.

\begin{theorem}
In a Markov chain defined as in~\Cref{sec:model}, a random walk starting from a sink of the transaction graph $G$ and progressing with transition probabilities defined by the transition matrix $P$ of~\Cref{eq:transmatrixdef} ends up being absorbed in an absorber node $v'\in V'$ with probability 1.
\end{theorem}
\begin{proof}
    We define the \emph{excess} function $b:V\rightarrow \mathbb{Z}$ as
    \begin{equation}
        b(u) = \sum_v a(v, u) - \sum_v a(u,v)\textrm{.}
    \end{equation}
    This means that $b(u)>0$ for sinks and $b(u)<0$ for sources of $G$.

    Let us consider the transaction graph extended with auxiliary nodes $V'$, using each edge backward. 
    Let $V^r$ denote the set of nodes $v$ reachable from a sink $s$ on a directed path in this graph, including the sinks as well. 
    
    We prove that each $w \in V^r$ is connected to at least one absorber $v'$ with a finite state sequence of non-zero probability. To see this, start a walk from a sink $s$ to $w$ and continue as follows. For each edge $(u,v)$ of the walk, set $a(u,v):=a(u,v)-1$. If this results in $a(u,v)=0$, delete the edge from the graph. Continue this until either reaching an absorber of the Markov chain (i.e. $u$ with $b(u)<0$), or a node where there are no more incoming edges.
    
    The walk from the sink $s$ to a absorber cannot progress infinitely since the maximum number of steps taken cannot exceed $\sum_{u,v} a(u,v)$. Further, apart from the starting and ending nodes, $b(u)$ remains identical for all intermediate nodes $u$. And finally, for a node $u$ with $b(u)\geq 0$, there must be at least one incoming edge where we can continue the walk. Since the only nodes with $b(u)<0$ are $u \in V'$, the walk must terminate in an absorber.
    
    Thus, from any node $v\in V^r$, there is a nonzero chance of at least $\epsilon_v$ of reaching an absorber in $k$ steps. Let $0<\epsilon = \min_v \epsilon_v$.
    Hence, the probability that a random walk from a sink does not reach an absorber in $k\cdot n$ steps is bounded by $(1-\epsilon)^n \rightarrow 0$.
    \end{proof}
    
     Note that the proof still allows separate components without sinks or sources to diverge, i.e., when $V^r\neq V\cup V'$. Such a component could be composed of, for example, a directed cycle with equal weights on all of the edges. These are not interesting to us since we are only interested in the absorbing entropies for sinks of the transaction graph $G$, so we can simply delete the relevant nodes.
\section{Graph Representation of Cryptocurrency Transactions} \label{sec:theory}
In this section, we show how one can arrive at a transaction graph from real-world cryptocurrency ledgers. We handle UTXO- and account-based cryptocurrencies separately and describe two ways of handling the temporality of transactions.

\subsection{Cryptocurrencies with the UTXO-model}\label{sec:txgraphinutxo}
In cryptocurrencies that use the unspent transaction output (UTXO) model, such as Bitcoin~\cite{nakamoto2008bitcoin} or Zcash~\cite{sasson2014zerocash}, money is represented as coins, also known as UTXOs. In a simplified form, each UTXO has a nominal value and defines a public key that can spend that output. Transactions consist of input and output coins. Each input references outputs of previous transactions that are not yet spent, and in return, transactions create new coins, referred to as output UTXOs. Transactions are identified by their cryptographic hash. A critical invariant enforced by cryptocurrency network participants is that the sum of all input values must be greater or equal to the sum of all output values in a transaction, i.e., it is impossible to print money out of thin air.

The UTXO model lends itself in a straightforward way to a weighted directed graph representation. Every transaction input and output gets a node in the graph. Every transaction is also represented by an identifier node in the graph. Each node corresponding to an input coin is connected to the transaction node with the appropriate value of the input coin as weight. Similarly, the transaction node is connected to each output node with a weight corresponding to the value of the output coin, \emph{ignoring transaction fees}. For an illustrative example, see~\Cref{fig:utxo_tx_graph}.

Since transaction outputs become the inputs of subsequent transactions, we obtain a transaction graph as desired. Once the transaction graph is obtained, we can transform the graph into an absorbing Markov chain as outlined in~\Cref{sec:tx_graph_to_markov_chain}. Note that the Markov chain transition probabilities only depend on the nominal values of the transaction inputs and outputs if we assume no background knowledge and apply the transition matrix described in~\Cref{eq:transmatrixdef}.

Multi-edges between nodes are, by definition, not possible in a UTXO-based representation. It is possible, however, to treat UTXO-based systems as account-based ones by merging nodes that belong to the same public key. We describe how we handle account-based systems in~\Cref{sec:txgraphinaccount}.

\subsection{Account-based Cryptocurrencies}\label{sec:txgraphinaccount}
Numerous cryptocurrencies use an account-based model, e.g., the most popular being Ethereum~\cite{wood2014ethereum}. In this model, each transaction moves value between a source and a destination account. Accounts can be smart contracts or externally owned accounts (EOA), i.e., when a key pair controls the account. We represent both EOAs and smart contracts by nodes in the transaction graph. For every transaction, we add an edge between the source and the destination with a weight equal to the corresponding transacted amount, again \emph{ignoring transaction fees}. Transforming the transaction graph to an absorbing Markov chain is done as in Section~\ref{sec:tx_graph_to_markov_chain}.

Multi-edges between nodes are possible in account-based systems. We handle these by merging all $u\rightarrow v$ transactions into a single $(u,v)$ edge, with the weight being the sum of the transferred amounts. Note, however, that this representation still allows for opposite edges, i.e., $(u,v)$ and $(v,u)$, to exist simultaneously.

\subsection{Modeling shielded pools}\label{sec:shielded_pools_modelling}
Certain cryptocurrencies apply privacy-enhancing techniques such as shielded pools, for example, Zcash~\cite{sasson2014zerocash}.

Zcash is a privacy-preserving cryptocurrency that applies zero-knowledge proofs to keep funds untraceable. 
It has two types of addresses: transparent and shielded. Transactions from, to, or between shielded addresses are hidden from the public using zero-knowledge proofs and, thus, cannot be traced. We model all shielded addresses as a single address representing the shielded pool as a whole. Each shielding transaction transfers money to this imaginary address, while each de-shielding transaction is represented by the pool address sending money to the recipient. We do not model transfers between shielded addresses, as these convey no useful information: they would materialize as self-loops of the pool address, leaving absorbing probabilities unaffected.

However, not acknowledging the specific nature of the shielded pool could result in an unreasonably low untraceability score.  Therefore, according to public knowledge, we model shielded pools not as a single node but as a collection of incoming transactions to the shielded pool. As a result, any output that is traced back to the pool ends up distributed between a large number of sources accurately representing the nature of the mixing characteristics of the transaction graph.

\subsection{Temporality of Transactions} \label{sec:temporaltxgraphs}

In a cryptocurrency system, each transaction has an implicitly assigned timestamp. Therefore, there is a total ordering for all transactions in a public ledger. One can either consider or disregard this temporal information when assessing traceability. Consequently, we consider two types of random walks on the cryptocurrency transaction graphs: \emph{stationary} and \emph{temporal}. We use the stationary version when evaluating a short time span, while we consider the temporal version more realistic for longer timespans.

\begin{figure}[tbhp]
    \centering
    \begin{subfigure}[b]{0.49\linewidth}
    \centering
        
            $\begin{bNiceMatrix}[first-row,first-col]
                ~ & \scriptstyle tx_1 & \scriptstyle a & \scriptstyle tx_3 \cr
                \scriptstyle {tx}_2& {\color{gray} 0} & 1 & {\color{gray} 0}\cr
                \scriptstyle tx_4& {\color{gray} 0} & 1 & {\color{gray} 0}\cr
                \scriptstyle a& \frac{\mathsf{a}}{\mathsf{a}+\mathsf{c}}&{\color{gray} 0} & \frac{\mathsf{c}}{\mathsf{a}+\mathsf{c}} \cr
            \end{bNiceMatrix}\hphantom{\scriptstyle tx_3}$
            
            \vspace{1em}
            \includegraphics[width=0.8\linewidth,clip]{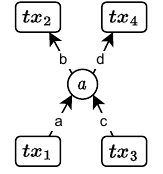}
        \caption{Stationary}\label{fig:temporalityExmplainer_stationary}
    \end{subfigure}
    \begin{subfigure}[b]{0.49\linewidth}
    \centering
    
        $\begin{bNiceMatrix}[first-row,first-col]
            ~ & \scriptstyle tx_1 & \scriptstyle a_1 & \scriptstyle tx_3 & \scriptstyle a_2 \cr
            \scriptstyle {tx}_2& {\color{gray} 0}& 1 & {\color{gray} 0} & {\color{gray} 0}	\cr
            \scriptstyle tx_4& {\color{gray} 0} & {\color{gray} 0} & {\color{gray} 0} & 1 \cr 
            \scriptstyle a_1& 1 &{\color{gray} 0}& {\color{gray} 0} & {\color{gray} 0} \cr
            \scriptstyle a_2&{\color{gray} 0} & \frac{\mathsf{a-b}}{\mathsf{a-b+c}} & \frac{\mathsf{c}}{\mathsf{a-b+c}}& {\color{gray} 0}
        \end{bNiceMatrix}\hphantom{\scriptstyle tx_3}$
        
            \vspace{1em}
        \includegraphics[width=0.8\linewidth,clip]{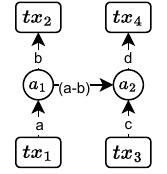}
        \caption{Temporal}\label{fig:temporalityExmplainer_temporal}
    \end{subfigure}

    \caption{Temporality of transactions. Four transactions, two incoming and two outgoing target the same address $a$. Transactions are created in the order $tx_1,tx_2,tx_3,tx_4$.  Weights on the edges represent the amount of the cryptocurrency transferred in the transactions. 
    The stationary and temporal way of representing them as a transaction graph is presented, with transition probability matrices also included in the figure. The stationary transaction graph disregards temporality information, while the temporal transaction graph incorporates it.  Note, as a result, the $tx_3\rightarrow tx_2$ path is blocked for a random walk in the temporal graph. 
    }
    \label{fig:temporalityExplainer}
\end{figure}

\paragraph{Stationary transaction graphs.}
In a stationary random walk, we disregard the timestamp information of transactions. Specifically, whenever we traverse back on the transaction graph, we allow the random walk to continue along any of the transactions, regardless of temporal order. In the example of Figure~\ref{fig:temporalityExplainer}, four transactions $tx_1,tx_2,tx_3,tx_4$ were included in the blockchain at times
$t_1,t_2,t_3,t_4$, with $t_1 < t_2< t_3< t_4$.
When a random walk starts at $\mathit{tx}_2$, we allow it to continue to either $\mathit{tx}_1$ or in $\mathit{tx}_3$, even though $\mathit{tx}_2$ happened earlier than $\mathit{tx}_3$. A stationary transaction graph captures the intuition that on a short enough time scale, the true source of an outgoing transaction of an address can be any of the incoming transactions, regardless of their temporality. This could more accurately represent the transfer of value between addresses in a short time frame. 
\paragraph{Temporal transaction graphs.}
In contrast to stationary transaction graphs, temporal transaction graphs take the timing of transactions into consideration to block older outgoing transactions from backtracking to newer incoming transactions. This is achieved by creating a new node representation $u_{i+1}$ for each time an account $u$ receives an incoming transaction and adding a $(u_i, u_{i+1})$ edge from the previous representation $u_i$ transferring the balance of $u_i$. This is reminiscent of the way UTXO-based systems work. Continuing with the example of Figure~\ref{fig:temporalityExplainer}, starting from $\mathit{tx}_2$, we can only backtrack to $\mathit{tx}_1$ during the random walk. It is not possible anymore to backtrack to $tx_3$, since $\mathit{tx}_2$ is older than $\mathit{tx}_3$, i.e., $t_2\leq t_3$.

In UTXO-based systems, the transaction graph is inherently temporal, so there is seemingly no difference between temporal and stationary representations. However, public key reuse makes this question relevant even in UTXO-based graphs, as the outputs of transactions can be treated as belonging to the same account when they share a public key. Though the practice of public key reuse is discouraged~\cite{conti2018survey}, it often happens in practice due to ease of use~\cite{moser2021resurrecting}.
\section{Applications} \label{sec:applications}
The flexibility of our proposed untraceability score allows for various applications. This section reviews a handful of these possible applications in detail. The two essential parameters of the proposed untraceability score $U_{\scriptscriptstyle\mathsf{S},G,\mathcal{A}}(v_i)$ are the considered source set $\mathsf{S}$ to which we wish to compute the absorbing probabilities, and the assumed adversarial knowledge $\mathcal{A}$ applied by the analyst who computes the untraceability score. We characterise three applications by their $\mathsf{S}$ and $\mathcal{A}$ parameters.

\paragraph{\textbf{Common-knowledge untraceability}} This variant of the untraceability wants to establish an upper bound on the untraceability of unspent coins with regard to coinbase UTXOs. The common-knowledge untraceability score solely assumes the knowledge of the transaction graph and disregards any additional knowledge that might decrease the untraceability score. In Section~\ref{sec:evaluation}, we thoroughly evaluate this type of untraceability score for several major cryptocurrency transaction graphs. The two parameters are:

\begin{itemize}
    \item $\mathsf{S}=\{v_i\vert v_i\in G \land\mathit{deg}_{\mathsf{in}}(v_i)=0\}$, i.e., all coinbase addresses.
    \item $\mathcal{A}=\mathbf{P}$, cf.~\Cref{eq:transmatrixdef}.
\end{itemize}

\paragraph{\textbf{A cryptocurrency exchange}} Cryptocurrency exchanges in most jurisdictions need to comply with anti-money laundering (AML) and know-your-customer (KYC) regulations. To that end, they collect, store, and process extensive information about the transaction graph. This additional knowledge allows them to cluster user addresses more effectively than only having access to the raw blockchain data. Most cryptocurrency exchanges do not allow money from illicit funds to be deposited. Let $\textsf{blackList}\subset V$ denote the set of nodes in the transaction graph that are blacklisted by the cryptocurrency exchange. A cryptocurrency exchange is interested in computing the probability distribution that a certain user fund residing at node $v\in V$ might contain funds from addresses in the \textsf{blackList}. Exchanges might apply the poison policy to transfer the taint between cryptocurrency transactions. The poison policy dictates that every transaction is tainted that has at least one dirty predecessor, no matter how many generations above~\cite{moser2014towards}. 

\begin{itemize}
    \item $\mathsf{S}=\textsf{blackList}$.
    \item $\mathcal{A}=\mathbf{P}^{\mathsf{pois}}$, cf.~\Cref{eq:transmatrixPoisdef}.
\end{itemize}

\paragraph{\textbf{Privacy-focused cryptocurrency wallet software}} Most privacy-cautious users want to break links from a set of addresses that they previously exposed through some means, e.g., social media. Let $\textsf{taintedUserAddresses}\subset V$  denote this subset of the transaction graph. A privacy-focused wallet must ensure that new user addresses are hardly traceable to previously leaked ones. Our untraceability metric can assist in this application scenario.

\begin{itemize}
    \item $\mathsf{S}=\textsf{taintedUserAddresses}$.
    \item $\mathcal{A}=\mathbf{P}$, cf.~\Cref{eq:transmatrixdef}.
\end{itemize}
\section{Evaluation} \label{sec:evaluation}
In this section, we compute and evaluate the untraceability metric proposed in~\Cref{sec:model} for various cryptocurrencies. To make the presented values comparable, we compute the traceability metric for each cryptocurrency for the same time interval between 2021 February $1$st and $8$th. This one-week time interval might seem like a short period, but the resulting transaction graphs already produce considerable transition graphs with millions of nodes and edges, cf.~\Cref{table:entropiesevaluation} and~\Cref{table:entropiesevaluation2}. These transaction graphs require significant computational effort to measure the proposed untraceability metric. Running time optimization for even larger transaction graphs are beyond the scope of the current paper. 

\begin{figure}[t]
    \centering
    \begin{subfigure}[t]{0.49\linewidth}
    \centering
        \includegraphics[width=\linewidth,clip]{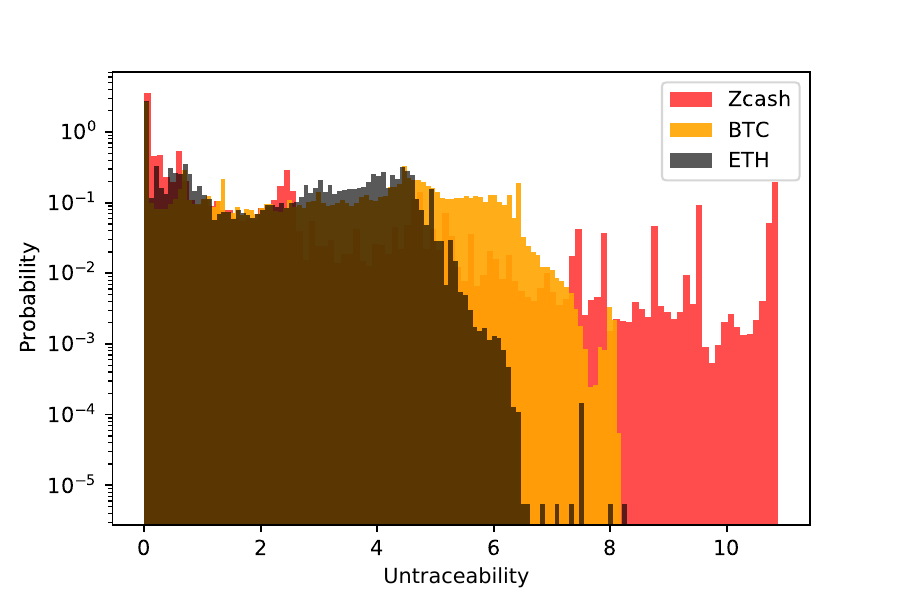}
    \end{subfigure}\hfill%
    \begin{subfigure}[t]{0.49\linewidth}
    \centering
        \includegraphics[width=\linewidth,clip]{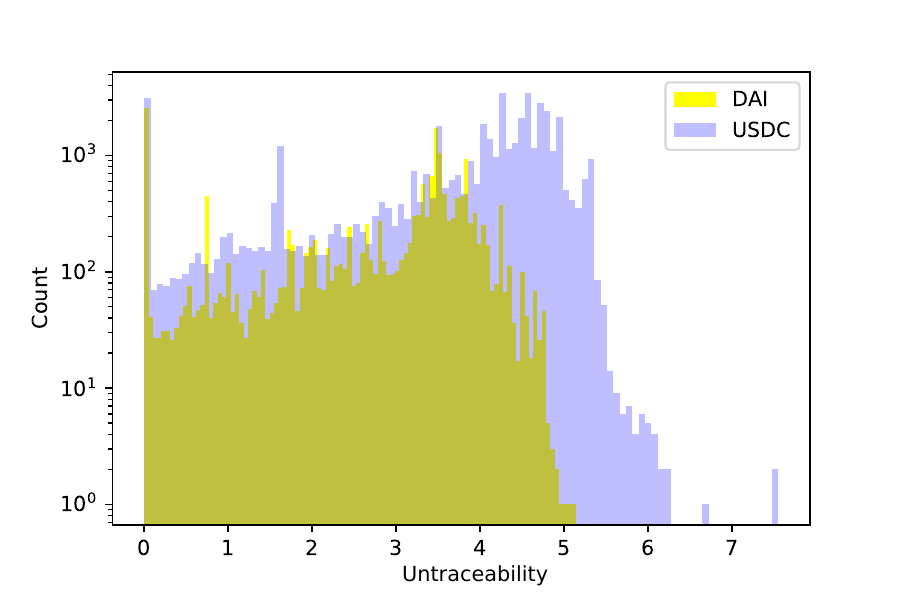}
    \end{subfigure}
    \caption{The distribution of untraceability of the unspent balances in the five studied cryptocurrencies (Bitcoin, Ethereum, Zcash (left), and DAI, USDC (right)) during the analyzed time period (2021 February 1-8).}
    \label{fig:entropiesComparisonStationary}
\end{figure}

We study both the stationary and the temporal transaction graph variants for each network, as described in~\Cref{sec:temporaltxgraphs}. We evaluate the traceability metric in the stationary variant in~\Cref{sec:stationaryevaluation} while also enclosing the evaluation of the temporary transaction graphs in~\Cref{sec:appendixtemporalevaluation}. For a summary of our results in the stationary variant, see~\Cref{table:entropiesevaluation} and~\Cref{fig:entropiesComparisonStationary}. The computations are carried out as described in~\Cref{sec:computational}. We set a value of  $\delta_{\text{threshold}}=0.001$ in~\Cref{eq:kthreshold}, which we think is a reasonable tradeoff between precision and efficiency when computing the untraceability of addresses.

In the evaluations presented in this Section, we treat Zcash as a special case. By taking a one-week snapshot of the transaction graph, most walks using de-shielding transactions are absorbed in the shielded pool, resulting in a small fungibility score. Since the shielding mechanism is a core component of Zcash, we instead modify the entropy values of~\Cref{def:fungibility} to include one more step after reaching the shielded pool, regardless of whether the step is during the considered one-week interval. We do this to account for the uncertainty introduced by shielding transactions that happened before the considered time interval. Formally, calculating the modified entropy $h'$ is done by observing that if
\begin{equation}
    h=-\sum_{i} \log_{2}(p_i) p_i\textrm{,}\quad\textrm{then}\quad
    h'=h-p_k\sum_{j}\log_{2}(q_j)q_j\textrm{,}
\end{equation}
where $h$ is the original untraceability score of the output according to absorbing probabilities $p_i$, the value $p_k$ is the probability of being absorbed at the node of the shielded pool, and $q_j$ is the distribution of in-edge values of the shielded pool from before the considered time interval. While the modified score is only an approximation, we feel it more accurately reflects the untraceability properties of Zcash. For reference, we also include the original results with unmodified entropy values.

\begin{table*}
\caption{Untraceability and the expected number of steps till absorption of unspent UTXOS (and addresses) in various cryptocurrencies. This table shows the results for the \emph{stationary transaction graph} variants; see~\Cref{sec:temporaltxgraphs}.}
\begin{center}
 \begin{tabular}{l c c c c c c c c c c} 
 \toprule
 &&&\multicolumn{4}{c}{Untraceability $(U(v))$}&&\multicolumn{3}{c}{Expected steps}\\
 \cline{4-7}\cline{9-11}
  & $\vert V \vert$ & $\vert E \vert$ &Mean&Median&Variance&Max&
  &Mean&Median&Variance\\
\midrule
 Bitcoin & $8598k$& $12801k$ & $3.96$ & $4.07$ &  $10.12$ & $11.82$ && $25.23$ & $23.56$ & $615.32$\\ 
 Zcash & $170k$ &$748k$ & $2.63$ & $0.89$ &  $15.36$ & $15.70$ && $52.66$ & $56.61$ & $1755.94$\\
 Ethereum & $3031k$& $5863k$ &  $3.45$ & $3.94$ & $5.97$ & $10.86$ && $7.27$ & $2.0$ & $80.51$\\
 DAI  & $46k$& $70k$ & $2.60$ & $3.30$ &  $1.98$ & $5.14$ && $26.39$ & $33.69$ & $414.49$\\
 USD Coin  & $146k$& $234k$ & $3.32$ & $4.10$ & $2.91$ & $8.12$&& $16.40$ & $18.35$ & $74.13$\\
 \bottomrule
\end{tabular}
\label{table:entropiesevaluation}
\end{center}
\end{table*}

\begin{table*}[ht!]
\caption{Untraceability and the expected number of steps till absorption of unspent UTXOs (and addresses) in various cryptocurrencies. This table shows the results for the \emph{temporal transaction graph} variants; see~\Cref{sec:temporaltxgraphs}.}
\begin{center}
 \begin{tabular}{l c c c c c c c c c c} 
 \toprule
 &&&\multicolumn{4}{c}{Untraceability $(U(v))$}&&\multicolumn{3}{c}{Expected steps}\\
 \cline{4-7}\cline{9-11}
  & $\vert V \vert$ & $\vert E \vert$ &Mean&Median&Variance&Max&
  &Mean&Median&Variance\\
\midrule
 Bitcoin & $11441k$& $13679k$ 
 & $3.97$ & $1.18$ &  $4.81$ & $10.34$
 && $20.64$ & $12.0$ & $632.21$\\ 
 Zcash & $946k$ &$1042k$
 & $1.02$ & $0.0$ &  $11.47$ & $15.94$
 && $533.96$ & $286.0$ & $394k$\\
 DAI  & $172k$& $252k$ 
 & $3.13$ & $3.23$ &  $4.00$ & $8.64$ 
 && $37.39$ & $23.80$ & $1489.19$\\
 USD Coin  & $508k$& $723k$ 
 & $3.62$ & $3.68$ & $3.83$ & $9.48$
 && $39.80$ & $27.87$ & $1577.23$\\
 \bottomrule
\end{tabular}
\label{table:entropiesevaluation2}
\end{center}
\end{table*}

\subsection{Stationary Transaction Graphs}\label{sec:stationaryevaluation}
\subsubsection{UTXO-based Cryptocurrencies}
\paragraph{\textbf{Bitcoin}}
We analyzed the Bitcoin transaction graph from block no.~$668\,548$ till no.~$669\,613$ and computed the fungibility of all $4\,237\,925$ unspent coins in this transaction graph segment. Note that many UTXOs have $0$ entropy. The mean entropy of coins is $3.96$ bits of entropy, while the median is $4.07$ bits. The maximum entropy of a coin in this transaction graph segment is $11.82$ bits. On average, it takes $25.23$ steps till an unspent coin is absorbed in a sink node, cf. Table~\ref{table:entropiesevaluation}. 

\begin{figure}[t]
    \centering
    \begin{subfigure}[t]{0.49\linewidth}
    \centering
        \includegraphics[width=\linewidth,clip]{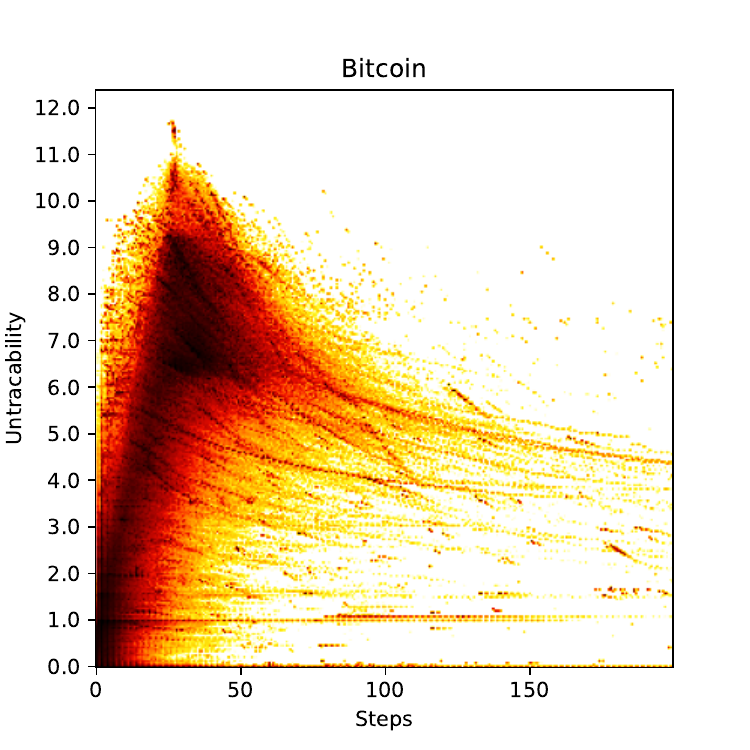}
    \end{subfigure}\hfill%
    \begin{subfigure}[t]{0.49\linewidth}
    \centering
        \includegraphics[width=\linewidth,clip]{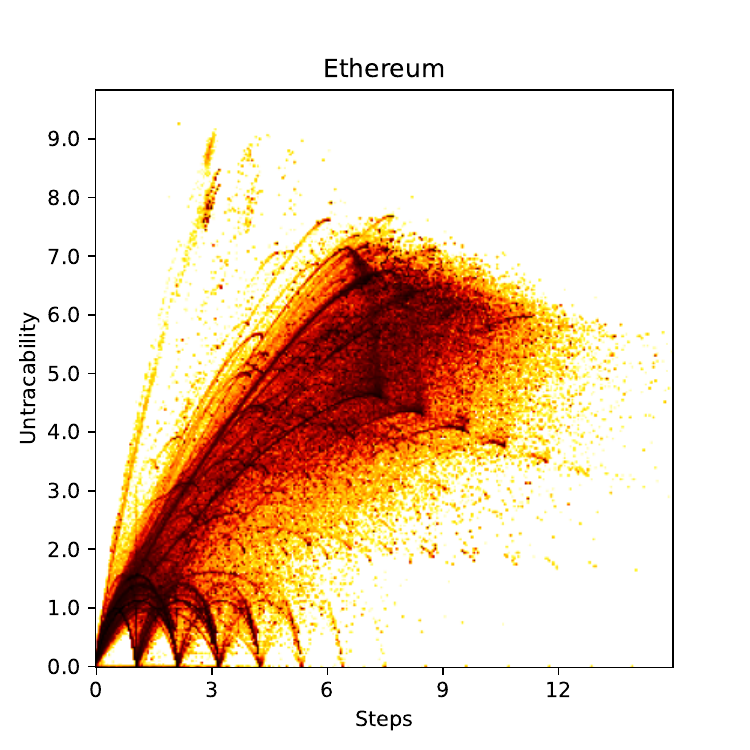}
    \end{subfigure}
    \caption{Scatter plots of untraceability and expected number of steps till absorption for Bitcoin (left) and Ethereum (right). Longer walks typically imply larger untraceability.}
    \label{fig:expectedStepsbyEntropyScatterPlot}
\end{figure}

\paragraph{\textbf{Zcash}}
The evaluated transaction graph segment corresponds to the transactions from block no.~$1\,131\,715$ until no.~$1\,139\,743$ with $94\,100$ transparent UTXOs. We observed the highest fungibility variance ($15.36$) among all studied cryptocurrencies. Perhaps unsurprisingly, we found the UTXO with the highest fungibility ($15.70$) in Zcash. Moreover, on average, we witnessed the longest expected time till absorption ($52.66$) in Zcash. 

\paragraph{\textbf{Zcash unmodified untraceability}}
For completeness, we enclose the untraceability results for Zcash when we did not attach any shielding ($t$-to-$z$) transactions issued before our chosen time interval. Naturally, this decreases the measured untraceability of unspent UTXOs. The unmodified Zcash graph in our  time window yielded an average untraceability $1.40$ with median $0.68$ and variance $1.99$. Once again, these results underscore the importance of the shielded pool, since the introduction of the shielding transactions in Section~\ref{sec:evaluation} increased untraceability on average by $0.88$.

\subsubsection{Account-based Cryptocurrencies}
\paragraph{\textbf{Ethereum.}}
We created a transaction graph from all transactions in the Ethereum blockchain that transferred non-zero amounts of Ether from block no.~$11\,766\,939$ till no.~$11\,812\,441$. Out of $1\,768\,192$ sink addresses, we observed a maximum untraceability of $10.86$ bits of entropy. We found the least number of expected steps till absorption  (mean $7.27$, median $2.0$) in the transaction graph, see~\Cref{table:entropiesevaluation}. We attribute these phenomena to the nature of the account-based model, where large initial balances lead to walks being absorbed early with high probability.
\paragraph{\textbf{ERC-20 token transaction graphs.}} We computed our untraceability measure for two stablecoins: USD Coin (USDC) and DAI. In the studied time interval these cryptoassets produced the smallest transaction graphs with $46$ and $146$ thousand addresses, respectively. Interestingly, both USDC and DAI random walks are absorbed after long random walks of expected length $26.39$ and $16.40$, respectively.

\subsection{Expected Number of Steps Till Absorption} \label{sec:appendixstationaryevaluation}
\begin{figure}[h]
    \centering
    \begin{subfigure}[t]{0.49\linewidth}
    \centering
        \includegraphics[width=\linewidth,clip]{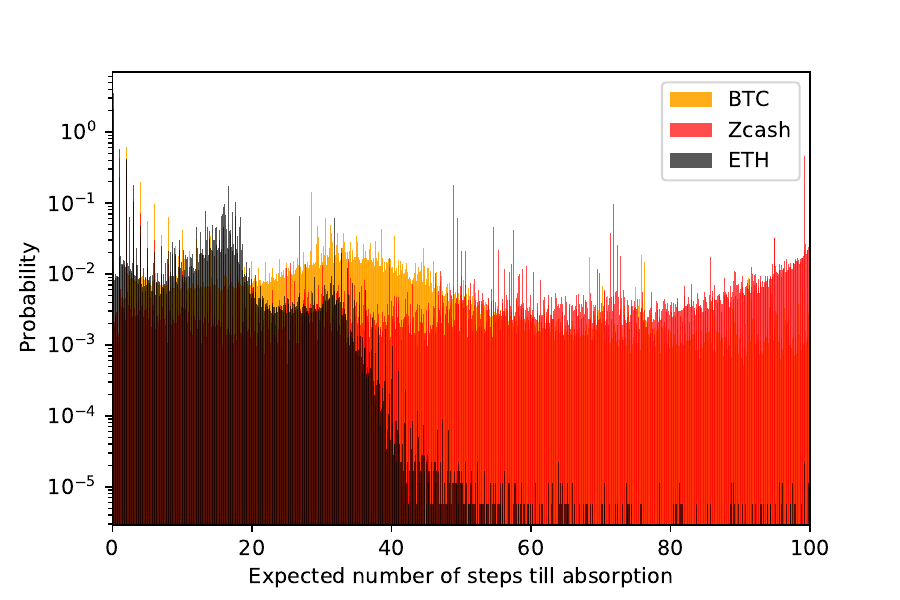}
    \end{subfigure}\hfill%
    \begin{subfigure}[t]{0.49\linewidth}
    \centering
        \includegraphics[width=\linewidth,clip]{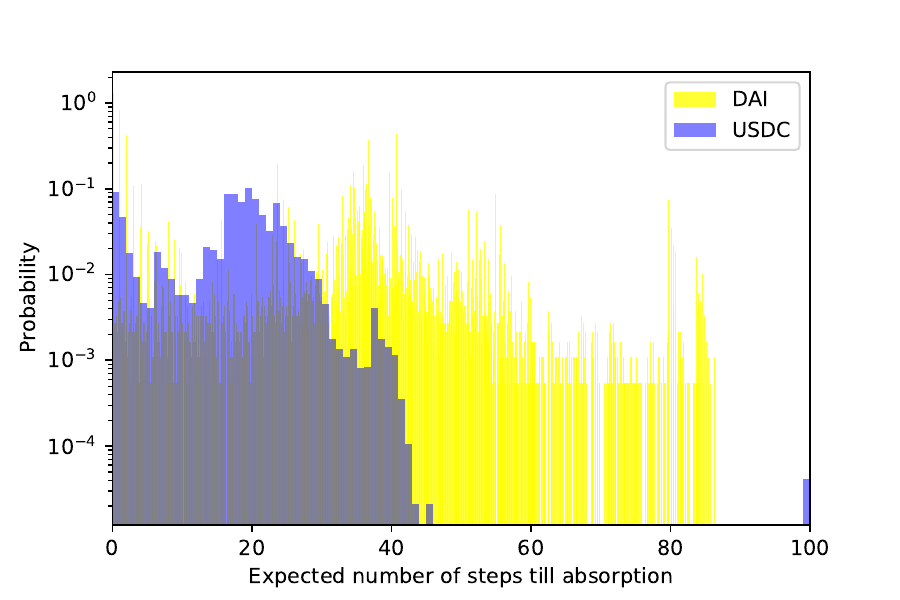}
    \end{subfigure}
    \caption{The expected number of steps till absorption for the studied cryptocurrencies (Bitcoin, Ethereum, Zcash (left), and DAI, USDC (right)).}
    \label{fig:expectedStepsHistograms}
\end{figure}

We computed the expected number of steps till absorption for every address with unspent balances in the studied transaction graph segments; see~\Cref{fig:expectedStepsHistograms}. It is striking in~\Cref{fig:expectedStepsHistograms} how quickly Ethereum random walks are absorbed, especially compared to Bitcoin and Zcash. Furthermore, it is fascinating that the DAI transaction graphs contain significantly longer random walks on average ($26.39$) than the USDC transaction graph ($16.40$). This holds despite the fact that the USDC graph ($\vert V \vert = 146$k, $\vert E\vert=234$k) is considerably larger than the DAI transaction graph ($\vert V \vert = 46$k, $\vert E\vert=70$k). 

\begin{figure}[h!]
    \centering
    \begin{subfigure}[t]{0.49\linewidth}
    \centering
        \includegraphics[width=\linewidth,clip]{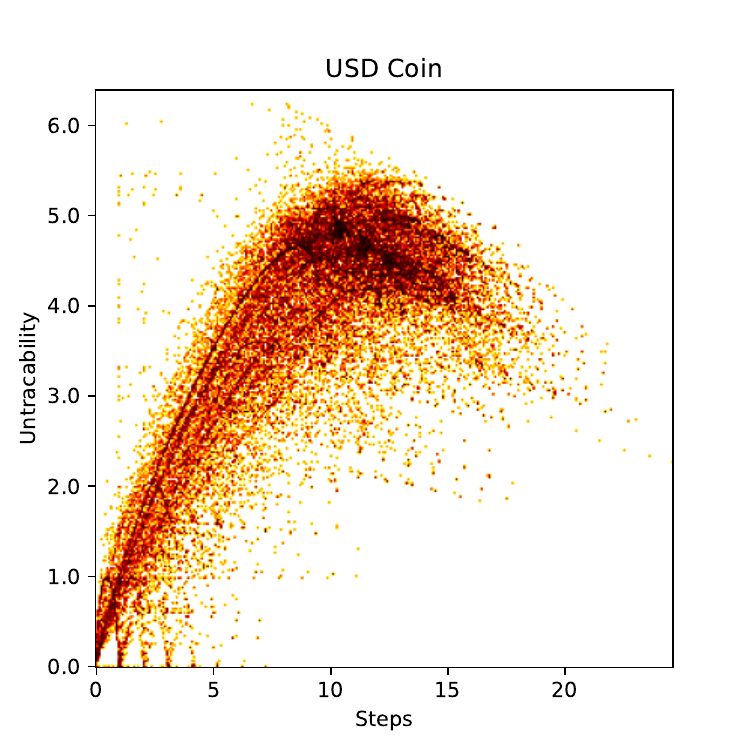}
        \caption{USD Coin.}
    \end{subfigure}\hfill%
    \begin{subfigure}[t]{0.49\linewidth}
    \centering
        \includegraphics[width=\linewidth,clip]{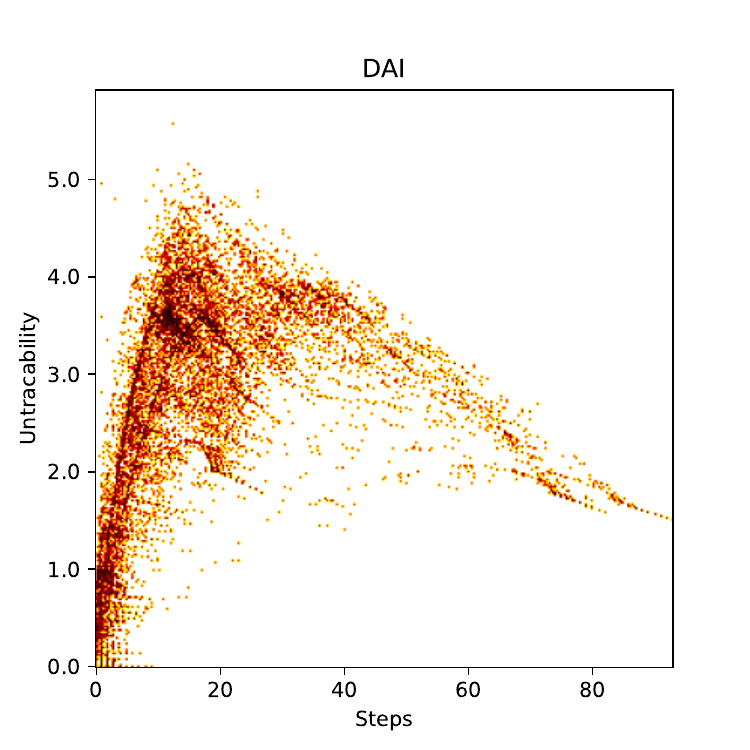}
        \caption{DAI.}
    \end{subfigure}
    \begin{subfigure}[t]{0.49\linewidth}
    \centering
        \includegraphics[width=\linewidth,clip]{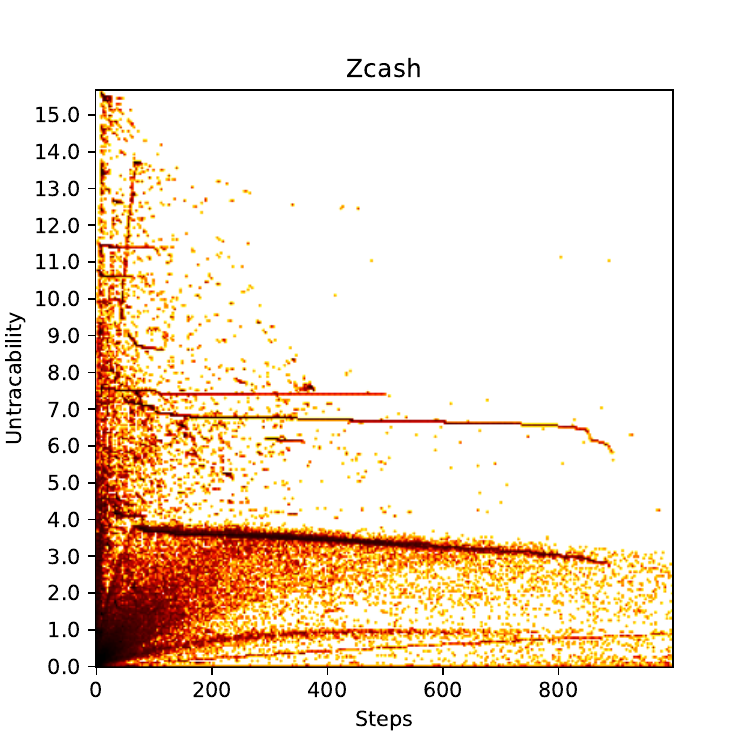}
        \caption{Zcash.}
    \end{subfigure}
    \caption{Contrasting entropies with the expected number of steps till absorption in the stationary transaction graphs.}
    \label{fig:entropyByStepsScatter}
\end{figure}

We provide the scatter plots of untraceability and the expected number of steps till absorption for USDC, DAI, and Zcash in~\Cref{fig:entropyByStepsScatter}. In USDC and DAI, there is a clear trend: the longer it is expected that a random walk is absorbed, the higher the untraceability it will achieve. However, larger random walks tend to decrease the achieved untraceability after a particular cutoff value, see~\Cref{fig:entropyByStepsScatter}.

On the other hand, Zcash exhibits completely different characteristics regarding untraceability and the expected lengths of random walks. First, Zcash does not need lengthy random walks to achieve high untraceability. We contribute this behavior to the shielded pool that effectively mixes together funds from different sources. Second, it seems that longer random walks decrease the untraceability score. Finally, we emphasize that Zcash achieves \emph{one magnitude longer} random walks on average than any other analyzed cryptocurrency in this work, cf.~\Cref{table:entropiesevaluation2}. 

\subsection{Temporal transaction graphs} \label{sec:appendixtemporalevaluation}
For each cryptocurrency discussed in the previous section, we analyze the temporal transaction graph as defined in~\Cref{sec:temporaltxgraphs}. We transform these temporal transactions graphs to absorbing Markov chains as discussed in~\Cref{sec:model}. We measure the untraceability and the expected number of steps till absorption for all unspent balances in the transaction graph segments. Due to the introduced auxiliary nodes, we observe larger graphs and Markov chains for all five considered cryptocurrencies. We believe these auxiliary nodes greatly contribute to \emph{reducing the untraceability} and the number of expected steps of all five cryptocurrencies. For the complete quantitative summary of results on temporal transaction graphs, see~\Cref{table:entropiesevaluation2}. 

\subsection{Discussion}

Generally, we can observe many different untraceability characteristics in the studied cryptocurrency networks. Some of the results are largely unsurprising, such as the relatively low mean expected number of steps in account-based currencies, as account balances can quickly absorb random walks. Similarly, a significant untraceability variance in Zcash is expected, as the two different kinds of Zcash addresses (i.e., shielded and transparent) exhibit significantly different untraceability characteristics.

On the other hand, the low mean and median untraceability of Zcash is surprising, as un-shielded transactions use the same UTXO technology as Bitcoin. We observe significant differences between DAI and USD Coin, which use virtually identical technology, i.e., the ERC-20 token standard. These statistics are affected by the number of transactions included in the one-week interval, as more transactions provide more mixing capability. Still, the differences lead us to believe that typical real-world usage of the systems also heavily affects observed untraceability besides technological choices. This is especially pronounced when observing that USDC has the largest median untraceability ($4.10$ bits) and comparable mean untraceability to Ethereum, even though both are account-based systems, and the latter saw more than $25$ times as much transaction activity over $20$ times as many nodes in the studied one-week time interval.

The two main cryptocurrency networks, Bitcoin and Ethereum, achieved comparable mean and median untraceability scores, but Bitcoin ultimately achieved a higher score in both. The similarity is remarkable since their technological choices result in different transaction graph structures, most evident when measuring the expected number of steps till absorption; see also~\Cref{fig:expectedStepsbyEntropyScatterPlot}.

\section{Limitations and Extensions} \label{sec:discussion}
\paragraph{\textbf{Limitations of our evaluation}}
When we take a one-week interval to study, the transaction graph's inputs are not particularly meaningful: they do not represent any known entities in general, unlike in some potential applications, see~\Cref{sec:applications}. Still, as our primary goal is to quantify the natural global mixing ability of each network, we believe that our experimental results demonstrate useful results towards this goal.

\begin{figure}[tbhp]
    \centering
\begin{subfigure}[b]{.5\textwidth}
    \centering
    \includegraphics[width=0.5\linewidth,clip]{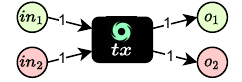}
    \caption{Mixing transaction example using Tornado Cash~\cite{pertsev2019tornado}.}\label{fig:tx_heuristics_a}
\end{subfigure}
\begin{subfigure}[b]{.22\textwidth}
    \centering
    $\begin{bNiceMatrix}[first-row,first-col]
    ~& \scriptstyle in_1 &  \scriptstyle in_2 & \scriptstyle tx & \scriptstyle {o}_1 & \scriptstyle {o}_2 \cr
     \scriptstyle {in_1} & {\color{gray} 0} & {\color{gray} 0} & 1 & {\color{gray} 0}	& {\color{gray} 0}\cr
     \scriptstyle {in_2} & {\color{gray} 0} & {\color{gray} 0} & 1 & {\color{gray} 0}	& {\color{gray} 0}\cr
     \scriptstyle {tx}&{\color{gray} 0} &{\color{gray} 0} & {\color{gray} 0} & 0.5 & 0.5	\cr
     \scriptstyle {o}_1 & {\color{gray} 0} &{\color{gray} 0}&  {\color{gray} 0}  & {\color{gray} 0}  & {\color{gray} 0}	\cr
     \scriptstyle {o}_2& {\color{gray} 0} & {\color{gray} 0}& {\color{gray} 0} & {\color{gray} 0}  & {\color{gray} 0} \nonumber
    \end{bNiceMatrix}$
    \caption{Transition matrix}\label{fig:tx_heuristics_b}
\end{subfigure}
\begin{subfigure}[b]{.22\textwidth}
    \centering
    $\begin{bNiceMatrix}[first-row]
    \scriptstyle in_1 & \scriptstyle in_2 & \scriptstyle tx & \scriptstyle {o}_1 &\scriptstyle {o}_2 \cr
    {\color{gray} 0} &{\color{gray} 0} & {\color{gray} 0} & 1 & {\color{gray} 0}\cr
    {\color{gray} 0} &{\color{gray} 0} & {\color{gray} 0} &  {\color{gray} 0} & 1\cr
    {\color{gray} 0} &{\color{gray} 0} &{\color{gray} 0} & {\color{gray} 0} & {\color{gray} 0}\cr
    {\color{gray} 0} &{\color{gray} 0} &{\color{gray} 0}&  {\color{gray} 0} & {\color{gray} 0}	\cr
    {\color{gray} 0} &{\color{gray} 0} & {\color{gray} 0}& {\color{gray} 0}  & {\color{gray} 0} \nonumber
    \end{bNiceMatrix}$
    \caption{Modified matrix}\label{fig:tx_heuristics_c}
\end{subfigure}
\caption{A toy example illustrating how to incorporate epistemic adversarial knowledge when computing our untraceability measure. A mixing transaction (a) with its corresponding transition matrix (b). If one can utilize additional information or heuristics to match deposit and withdraw addresses belonging to the same entity, then the transition matrix can be modified to reflect this knowledge (c).
}
    \label{fig:tx_heuristics}
\end{figure}

In our evaluation, certain services increase the untraceability score, even if the actual flow of money through the service can be unambiguously inferred from public knowledge. Consider star-shape transaction graphs, where numerous accounts send and receive coins from a single address, e.g., Satoshi dice\footnote{See:~\url{https://satoshidice.com/}.}, decentralized exchanges, e.g., Uniswap~\cite{adams2021uniswap}, etc. One could refine the transition matrix, see~\Cref{fig:tx_heuristics} for an illustrative example, by connecting the in- and out-flow of money through these services, using the additional knowledge of which input address corresponds to which output address.

\paragraph{\textbf{Obtaining a measure of anonymity}}

So far, we have exclusively focused on creating a formal model to calculate the untraceability in cryptocurrency transaction graphs. Intuitively, one can consider our proposed metric as a way to compute an upper bound on the uncertainty about the origins of unspent coins. However, in many occasions, one could apply heuristics~\cite{beres2021blockchain,moser2021resurrecting,wu2022tutela}, private knowledge, or additional information stored in the blockchain or obtained from other sources to reduce the uncertainty about the origins of coins. This knowledge could be incorporated into the definition of the transition probabilities, cf.~\Cref{eq:transmatrixdef}. Since these heuristics and additional information sources are strongly dependent on the adversaries, this notion could measure \emph{the particular adversary's uncertainty} in assessing the origins of certain coins. Hence, this modified metric could be considered as an anonymity metric. Consider the transaction graph segment in~\Cref{fig:tx_heuristics_a}, depicting a mixing transaction, e.g., Tornado Cash~\cite{pertsev2019tornado}, with two deposit and two withdraw addresses. The two withdraws are equally likely to come from either deposit, resulting in the transition matrix presented in Figure~\ref{fig:tx_heuristics_b}.

However, if one can infer that $\mathsf{in}_1$ and $\mathsf{in}_2$ belong to the same entities as $o_1$ and $o_2$ respectively, for example by pairing heuristics based on certain usage characteristics described in~\cite{beres2021blockchain}, then the transition matrix can be modified to reflect this knowledge, see~\Cref{fig:tx_heuristics_c}.

\section{Related Work} \label{sec:relatedwork}
\begin{table*}[ht!]
\centering
\resizebox{\textwidth}{!}{ 
 \setlength{\tabcolsep}{3pt}
 \setlength{\belowbottomsep}{6pt}
 \begin{tabular}{lcccccc} 
 \toprule
  \textbf{Approach} & Range & Fine-grained &Decay&Tamper-resistant & Subjective &Efficiently Computable \\
\midrule 
$k$-anonymity~\cite{sweeney2002k}                    & \cmark             &\xmark    &\cmark &\cmark&\cmark &\cmark              \\
Degree of anonymity~\cite{diaz2002towards,serjantov2002towards}& \cmark             &\cmark    &\cmark &\cmark&\cmark &\xmark\\
System's anonymity~\cite{edman2007combinatorial,gierlichs2008revisiting}& \cmark             &\cmark    &\cmark &\cmark&\cmark &\xmark\\
Differential privacy~\cite{dwork2006calibrating}& \cmark             &\xmark    &\xmark &\cmark&\xmark &\cmark\\
Boltzmann score~\cite{laurentmt2017boltzmann}& \cmark             &\xmark    &\cmark &\cmark&\cmark &\xmark\\
Wicht et al.~\cite{wicht2023transaction}& \cmark             &\xmark    &\cmark &\cmark&\cmark &\cmark\\\midrule
Our approach                                    & \cmark                 & \cmark    &\cmark & \cmark  &\cmark & \cmark          \\ 
 \bottomrule
\end{tabular}
}
\caption{Comparison of major approaches for measuring cryptocurrencies' (un)traceability or related notions applicable to transaction graphs, see~\Cref{sec:relatedwork}. For a list of features, see~\Cref{sec:desiderata}. }
\label{table:approaches_qualitative_comparison}
\end{table*}

There are several previous attempts in the literature to \emph{quantify} traceability and related notions, such as anonymity, privacy, or fungibility in the context of cryptocurrencies. A commonly used privacy metric called $k$-anonymity, was introduced by Sweeney~\cite{sweeney2002k}. The notion of $k$-anonymity is sometimes also referred to as plausible deniability. It is extensively used to argue about the anonymity guarantees of a \emph{single cryptocurrency transaction}, e.g., CoinJoin transactions~\cite{maxwell2013coinjoin} or cryptocurrency mixers~\cite{beres2021blockchain}. However, $k$-anonymity does not generalize to multiple transactions meaningfully. Additionally, $k$-anonymity is inherently incapable of capturing the probabilistic nature of anonymity, i.e., the probability of each object in the anonymity set is uniform. The probabilistic nature of anonymity is captured by an information-theoretic approach proposed by Diaz et al.~\cite{diaz2002towards} and Serjantov et al. ~\cite{serjantov2002towards}. In their landmark works, the adversary outputs a probability distribution of guessing a target user in the anonymity set. The Shannon entropy of this distribution is considered as the degree of anonymity, i.e., the adversary's uncertainty about the target. Our work can be interpreted as generalising this entropic approach to the blockchain setting.

Boltzman score was proposed by LaurentMT\footnote{See: Boltzman score: \url{https://gist.github.com/LaurentMT/e758767ca4038ac40aaf}.} to quantify the anonymity guarantees of single Bitcoin transactions~\cite{laurentmt2017boltzmann}. The Boltzman score computes all the possible groups of the transaction's outputs that can be obtained by grouping the transaction's inputs. The number of input-output assignments is then divided by the possible number of input-output assignments. This notion is only computable for small Bitcoin transactions, i.e., with a small number of inputs or outputs, because computing this score reduces to the NP-complete subset sum problem~\cite{kleinberg2006algorithm}. In contrast, in this work, we are interested in defining a privacy notion that is efficiently computable even for large real-world transaction graphs. 

Very recently, Wicht et al. ~\cite{wicht2023transaction} introduced an untraceability and unlinkability metric for transaction graphs. They introduce a novel transaction graph model for cryptocurrencies, in which they model the transaction graph as a bipartite graph where one class of nodes consists of the inputs to transactions, while the other contains the transaction outputs. At a high level, their untraceability notion could be obtained by computing matchings on their own transaction graph model. Their notion can be seen as a generalization of the k-anonymity notion to the blockchain/transaction graph setting. Unfortunately, this metric is unable to express the probabilistic nature of the observed privacy notions. More importantly, they did not implement and evaluate their proposed metric on cryptocurrency transaction graphs. Hence, unfortunately, we cannot compare our empirically evaluated untraceability metric with theirs.

Several \emph{qualitative} frameworks grasp various notions of privacy and anonymity in anonymous communication systems by Backes et al.~\cite{backes2013anoa} and in cryptocurrency systems by Amarasinghe~\cite{amarasinghe2021cryptographic}. They define numerous flavours of unlinkability, indistinguishability, and anonymity in their corresponding settings. A significant limitation of these works is that they do not extend to multiple subsequent transactions, i.e., they solely analyze their privacy and anonymity notions for isolated transactions. Our main goal is to measure these notions for entire transaction graphs obtained from public ledgers.

To the best of our knowledge, no efficient metric has been proposed in the literature to measure the (un)traceability of money. A theory of fungibility was developed in~\cite{shorish2021practical}. However, it does not provide a way to measure fungibility in cryptocurrencies. The closest related work was done by Pontiveros et al.~\cite{pontiveros2019mint}, where the authors proposed a notion called mint centrality. Mint centrality measures for a coin $u$, how many other coinbase coins (minted coins) can be traced back to $u$ via transactions. Our metric is a strict generalization of mint centrality as we also measure how much a particular coinbase coin has contributed to a specific coin's fungibility throughout possibly many transactions.
\section{Conclusion and Future Work} \label{sec:conclusion}
In this work, we proposed a mathematical model based on absorbing Markov chains and Shannon entropy for a quantitative measure assessing the untraceability of cryptocurrencies. Additionally, we extensively evaluated the introduced untraceability measure in the same one-week time period on multiple UTXO- and account-based cryptocurrencies. Finally, we discussed how one can transform this untraceability measure into a degree of anonymity, incorporating exogenous adversarial knowledge into the computation of the untraceability measure. Future work entails studying and comparing the quantitative untraceability benefits of additional privacy-enhancing techniques and cryptocurrency designs, e.g., ring signatures applied in Monero~\cite{van2013cryptonote}, the various Coinjoin protocols of Bitcoin~\cite{ficsor2021wabisabi,ghesmati2022sok}, stealth addresses deployed on Ethereum~\cite{courtois2017stealth,wahrstatter2024basesap}, confidential transactions~\cite{poelstra2018confidential,bunz2020zether} or the Darksend design of Dash~\cite{duffield2015dash}.
\ifarxiv
\paragraph{Acknowledgements.}
This work has been partially funded by the European Union project RRF-2.3.1-21-2022-00004 within the framework of the Artificial Intelligence National Laboratory.
We are thankful to Blockchair\footnote{See:~\url{https://blockchair.com}.} for being a generous data provider for our on-chain statistics. We are grateful to András A. Benczúr, Balázs Pejó, Dániel Fehér, Ferenc Béres, and Zsombor Jancsó for insightful discussions. We are indebted to Ádám Ficsór and Yuval Kogman for raising the quantification of fungibility and privacy as a fascinating research question.
\fi

\bibliographystyle{plain}
\bibliography{references}

\appendix
\section{Change of measured untraceability with walk length}
\begin{figure}[h]
    \centering
    \begin{subfigure}[t]{0.49\linewidth}
    \centering
        \includegraphics[width=\linewidth,clip]{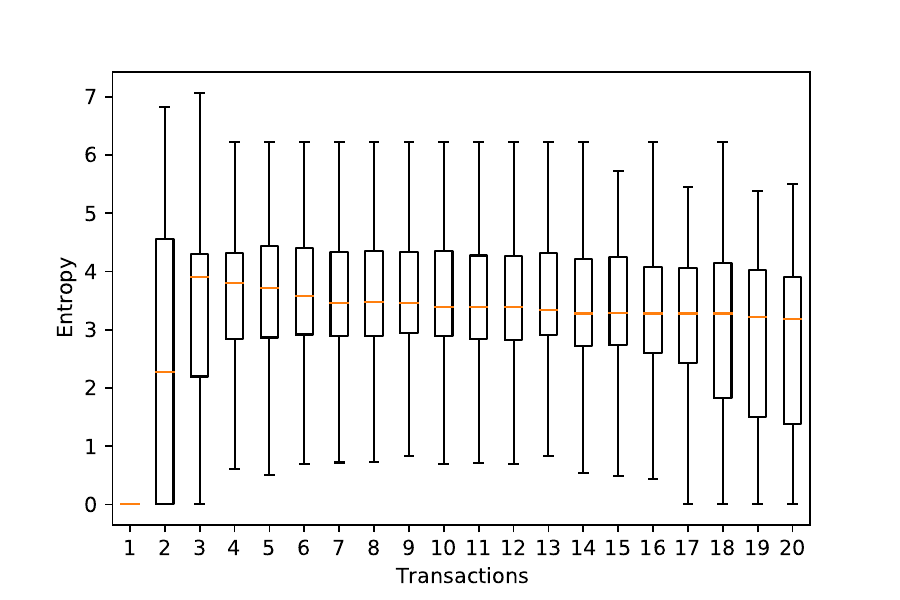}
        \caption{Ethereum.}
    \end{subfigure}\hfill%
    \begin{subfigure}[t]{0.49\linewidth}
    \centering
        \includegraphics[width=\linewidth,clip]{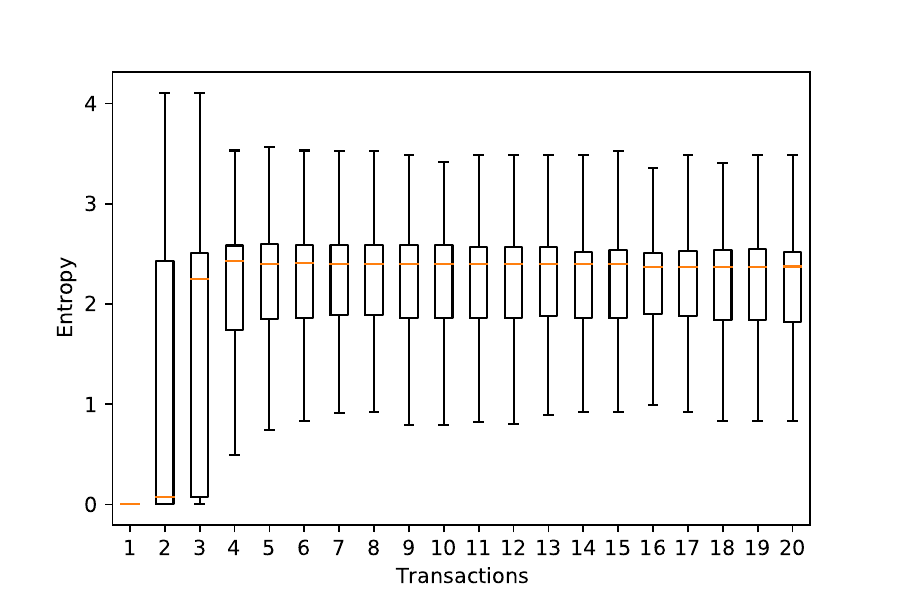}
        \caption{DAI.}
    \end{subfigure}
    \caption{The change in untraceability through time.}
    \label{fig:entropiesThroughTime}
\end{figure}

Last but not least, we analyzed the effect of time on untraceability. More precisely, starting from every absorber node, e.g., coinbase addresses, we computed the achieved untraceability. Afterwards, we proceeded in a random walk from the absorber node and recorded how untraceability changes, see Figure~\ref{fig:entropiesThroughTime}. This experiment emulates the effect of subsequent transactions on untraceability. As Figure~\ref{fig:entropiesThroughTime} shows, untraceability rapidly increases after the first handful of transactions. However, it quickly saturates, and further transactions seem to decrease untraceability.

\end{document}